\documentclass[aps,prb,longbibliography,showpacs,floatfix,superscriptaddress,twocolumn]{revtex4-1}

\usepackage{graphicx,color}
\usepackage{amsfonts}
\usepackage[figuresright]{rotating}  
\usepackage{amssymb}
\usepackage{amsmath}
\usepackage{mathtools}
\usepackage{psfrag}
\usepackage{subfigure}
\usepackage{multirow}
\usepackage{tabularx}
\usepackage{textcomp}
\usepackage{units}
\usepackage{lipsum}
\usepackage{soul}
\usepackage{titlesec}
\usepackage{times}
\usepackage{hyperref}
\usepackage{enumitem}
\usepackage{physics}

\usepackage{cleveref}
\usepackage{braket}
\usepackage{mathrsfs}

\DeclareMathAlphabet\mathbfcal{OMS}{cmsy}{b}{n}

\titlespacing*{\section}
{0pt}{1ex plus .25ex}{1ex plus 1ex}
\titlespacing*{\subsection}
{0pt}{1ex plus .25ex}{1ex plus 1ex}
\titlespacing*{\subsubsection}
{0pt}{1ex plus .25ex}{1ex plus 1ex}

\titleformat{\section}
 {\fontsize{10}{15}\bfseries }
  {\thesection}
  {1em}
  {}

\titleformat{\subsection}
  {\bfseries }
  {\thesubsection}
  {2em}
  {}

\titleformat{\subsubsection}
  {\itshape}
  {\thesubsubsection}
  {2em}
  {}

\def\tr{\mbox{tr}}

\def\beq{\begin{eqnarray}}
\def\eeq{\end{eqnarray}}

\renewcommand{\v}[1]{\ensuremath{\mathbf{#1}}} 
 
\newcommand{\bk}{\boldsymbol{k}} 
\newcommand{\mc}[1]{\mathcal{ #1}} 

\makeatletter

\titleclass{\subsubsubsection}{straight}[\subsection]

\newcounter{subsubsubsection}[subsubsection]
\renewcommand\thesubsubsubsection{\thesubsubsection.\arabic{subsubsubsection}}

\titleformat{\subsubsubsection}
   {\normalfont\normalsize\itshape \centering}{\thesubsubsubsection}{1em}{}
\titlespacing*{\subsubsubsection}
{0pt}{1ex plus .3ex}{1ex plus .3ex}

\makeatletter
\renewcommand\paragraph{\@startsection{paragraph}{5}{\z@}%
  {3.25ex \@plus1ex \@minus.2ex}%
  {-1em}%
  {\normalfont\normalsize}}
\renewcommand\subparagraph{\@startsection{subparagraph}{6}{\parindent}%
  {3.25ex \@plus1ex \@minus .2ex}%
  {-1em}%
  {\normalfont\normalsize}}
\def\toclevel@subsubsubsection{4}
\def\toclevel@paragraph{5}
\def\toclevel@paragraph{6}
\def\l@subsubsubsection{\@dottedtocline{4}{7em}{4em}}
\def\l@paragraph{\@dottedtocline{5}{10em}{5em}}
\def\l@subparagraph{\@dottedtocline{6}{14em}{6em}}
\makeatother

\begin{document}
\title{Effective field theory of the quantum skyrmion Hall effect}

\author{Vinay Patil}

\affiliation{Max Planck Institute for the Physics of Complex Systems, Nöthnitzer Strasse 38, 01187 Dresden, Germany}
\affiliation{Max Planck Institute for Chemical Physics of Solids, Nöthnitzer Strasse 40, 01187 Dresden, Germany}

\author{Rafael Flores-Calderón}

\affiliation{Max Planck Institute for the Physics of Complex Systems, Nöthnitzer Strasse 38, 01187 Dresden, Germany}
\affiliation{Max Planck Institute for Chemical Physics of Solids, Nöthnitzer Strasse 40, 01187 Dresden, Germany}

\author{Ashley M. Cook}
\affiliation{Max Planck Institute for the Physics of Complex Systems, Nöthnitzer Strasse 38, 01187 Dresden, Germany}
\affiliation{Max Planck Institute for Chemical Physics of Solids, Nöthnitzer Strasse 40, 01187 Dresden, Germany}
\begin{abstract}
Motivated by phenomenology of myriad recently-identified topologically non-trivial phases of matter, we introduce effective field theories (EFTs) for the quantum skyrmion Hall effect (QSkHE). We employ a single, unifying generalisation for this purpose: in essence, a lowest Landau level projection defining a non-commutative, fuzzy sphere with position coordinates proportional to SU(2) generators of matrix representation size $N\times N$, may host an intrinsically 2+1 dimensional, topologically non-trivial many-body state for small $N$ as well as large $N$. That is, isospin degrees of freedom associated with a matrix Lie algebra with $N \times N$ generators potentially encode some finite number of spatial dimensions for $N\ge 2$, a regime in which isospin has previously been treated as a label. This statement extends to more general $p$-branes subjected to severe fuzzification as well as membranes. As a consequence of this generalisation,
systems with $d$ Cartesian spatial coordinates and isospin degrees of freedom encoding an additional $\delta$ fuzzy coset space coordinates can realise topologically non-trivial states of intrinsic dimensionality up to $d$+$\delta$+1. We therefore identify gauge theories with extra fuzzy dimensions generalised to retain dependence upon gauge fields over fuzzy coset spaces even for severe fuzzification (small $N$), as EFTs for the QSkHE. We furthermore generalise these EFTs to space manifolds with local product structure exploiting the dimensional hierarchy of (fuzzy) spheres. For this purpose, we introduce methods of anisotropic fuzzification and propose formulating topological invariants on fuzzy coset spaces as artifacts of projecting matrix Lie algebras to occupied subspaces. Importantly, we focus on phenomenology indicating the 2+1 D SU(2) gauge theory should be generalised using this machinery, and serves as a minimal EFT of the QSkHE. 
\end{abstract}
\maketitle
\tableofcontents  

\section{Introduction}
Spin angular momentum is a fundamental concept of quantum mechanics, introduced by Pauli~\cite{pauli_uber_1925, Pauli1927ZurQD}, which is elegantly encoded in matrix mechanics~\cite{1925ZPhy...34..858B, 1926ZPhy...35..557B} dating back to study of the quantum harmonic oscillator by Heisenberg \cite{1925ZPhy...33..879H}. This formulation is now broadly applied across modern physics. It is essential to condensed matter physics, for instance, where some of the most challenging problems relate to strongly-correlated phases of matter exhibiting complex magnetic order intertwined with other phenomena~\cite{sachdev2018topological, PhysRevX.8.021048, PhysRevLett.119.227002, sachdev2016novel}. Spin encoded in matrix representations is central to quantum information physics, encoding myriad quantum states and associated gate operations~\cite{cirac1995, clarke_superconducting_2008, nielsen2010quantum, arute2019quantum, Preskill2018quantumcomputingin}. These methods are also essential to high-energy physics, where spin provides structure to the Standard Model in labeling and relating myriad subatomic particles\cite{1939AnMat..40..149W, Weinberg:1995mt}, for instance.

At the frontiers of modern physics, however, interpretation and treatment of spin angular momentum in quantum mechanics is still being expanded to confront the challenges of unification of the four forces. Theories of quantum gravity at the forefront of high-energy physics \cite{Georgi:1974sy, Feynman:1963ax, Donoghue:1995cz}, in particular, tackle scenarios in which general relativity is confronted by quantum physics, such as at the event horizon of a black hole \cite{Hawking:1975vcx, tHooft:1984kcu, RevModPhys.88.015002}. The dimensionality of space-time is essential to these efforts, such as via the Anti de Sitter, conformal field theory correspondence\cite{Witten:1998qj,Witten:1998zw}, or AdS-CFT. Considerable work of high-energy physics has explored, as well, consequences of varying the number of space-time dimensions\cite{Kaluza:1921tu, Klein:1926tv, Arkani-Hamed:1998jmv,PhysRevLett.83.3370,Sean_M_Carroll_2009}, with many theories containing more than the 3+1 dimensions (3 spatial dimensions and 1 time dimension) of known space-time as part of efforts to make general relativity consistent with quantum field theory. Spin angular momenta associated with matrix Lie algebras have emerged as useful tools in encoding spatial dimensions\cite{Aschieri:2003vy, Aschieri:2004vh, gavriil2015higher, aschieri2007, Kapetanakis:1992hf, balachandran2006, hasebe2024, Hasebe:2010vp, Hasebe:2014nia, Chatzistavrakidis:2010tq}. This approach also encodes non-commutativity that appears, for instance, due to the presence of external fields or restrictions to subsets of Hilbert space~\cite{susskind2001_qh, aschieri2007, SimeonHellerman_2001}.

While spin angular momentum is used in these effective theories to encode spatial dimensions to great effect when the matrix representation size of the Lie algebra generators is large (generators are represented by $N\times N$ matrices, with $N$ large)~\cite{matusis2000, SimeonHellerman_2001, Zhu:2022gjc, hu2023}, spin angular momentum is treated as a quantum number---effectively as just a label---for quantum states when $N$ is small, such as in study of quantum spin liquids and other strongly-correlated phases of matter~\cite{ANDERSON1973153, ANDERSON87, BASKARAN1987973, PhysRevLett.59.2095, PhysRevB.35.8865, PhysRevLett.61.2376, PhysRevB.37.3774, PhysRevLett.66.1773, PhysRevB.40.7387, PhysRevB.44.2664, PhysRevB.62.7850, PhysRevLett.86.292, wegner1971duality, RevModPhys.51.659, PhysRevLett.86.1881, KITAEV20032, KITAEV20062}. In this small $N$ regime, spin is termed isospin specifically to convey it is an internal degree of freedom (DOF). It is also termed (pseudo)spin (pspin) in the condensed matter literature, to convey that it could be associated with the physical spin or some other DOF, such as a layer or valley DOF. 

One can then naively ask, are these two treatments of spin angular momentum in quantum systems---at large $N$ and at small $N$---consistent? If so, how exactly does the cross-over from interpretation of spin as encoding spatial dimensions at large $N$ to interpretation of spin as a label at small $N$ occur? This question is deeply relevant to the quantum skyrmion Hall effect (QSkHE)~\cite{qskhe} and related topologically non-trivial phases of matter~\cite{cook_multiplicative_2022, cook2023, cookFST2023} recently-identified in condensed matter systems, and we address it as part of introducing the effective field theory (EFT) of the QSkHE. We find that spin angular momentum encoded in matrix Lie algebras may encode intrinsically $\delta+1$ ($\delta$ spatial dimensions and 1 time dimension) dimensional topologically non-trivial phases of matter, even for small non-trivial $N$ in the regime where spin has previously been treated as a label~\cite{qi2008TRIFT, PhysRevLett.73.874, EOM00, PhysRevB.49.17208, PhysRevB.51.14725, KASNER1995289, PhysRevB.54.R2331, PhysRevLett.76.3204, V_N_Nicopoulos_1995}. A system with $D$ Cartesian space coordinates and $\delta$ dimensions encoded in spin angular momentum can then potentially realise states of intrinsic dimensionality $D+\delta+1$. This explains many phenomena of the QSkHE, where topology associated with isospin degrees of freedom (DOFs) is also associated with bulk-boundary correspondence and response signatures of higher-dimensional-than-expected topologically non-trivial phases of matter. It furthermore corresponds to a generalisation of the quantum Hall effect (QHE) framework~\cite{PhysRevLett.45.494, PhysRevLett.48.1559, laughlin1981, kallin1984excitations, halperin1993theory, halperin1982quantized, halperin1984statistics, wen1991gapless, PhysRevLett.50.1395, PhysRevLett.62.82, PhysRevLett.63.199, PhysRevLett.64.1313, PhysRevB.47.16419, PhysRevB.78.195424, Ryu_2010, schnyder2008}, in that intrinsically $\delta+1$-dimensional many-body states encoded in spin angular momentum, even for small $N$, can play the role that charged particles play in the QHE.

Based on this generalisation to interpretation of spin as encoding some number of spatial dimensions, even for small matrix representation size of the associated matrix Lie algebra generators, we introduce an EFT for the QSkHE consistent with a growing number of related works~\cite{cook_multiplicative_2022, cook2023, cookFST2023, calderon_skyrm, calderon2023_fst, pal_multsemimetal, pal2023multiplicative, pal2024_fstwsm, pacholski2024, ay2023, liu2023skyrmsemimetals}, by generalising  quantum field theories with extra fuzzy dimensions~\cite{aschieri2007, Aschieri:2004vh, Aschieri:2003vy}, defined over manifolds with global product structure, to those with local product structure, and retaining dependence on fuzzy gauge fields in the small $N$ regime. To do so, we introduce a method of anisotropic fuzzification to derive a minimal EFT for the QSkHE starting from the 4+1 D SU(2) gauge theory of the 4+1 D QHE~\cite{zhanghu2001, bernevig6Dfieldtheory,Demler:1998pm,RevModPhys.76.909}. We also propose methods to compute quantised topological invariants over fuzzy, non-commutative spaces from structure constants of matrix Lie algebras for observable operators \textit{projected to occupied subspaces}, which we term \textit{structure factors}, generalising the notion of a structure constant of a matrix Lie algebra. In analogy to a non-trivial Chern number occurring as an artifact of projection to an occupied subspace\cite{PhysRevB.85.241308,parameswaran2013fractional, Estienne:2012bi,Neupert:2012gk}, we similarly identify non-trivial artifacts of these projections within the Lie algebra structure factor. We are furthermore able to revisit previously-unexplained observations~\cite{ma_unexpected_2015} in experiments on HgTe quantum wells (QWs)~\cite{QSHI-HgTe-Theory}, finding they are consistent with this new interpretation of spin angular momentum in quantum mechanics, and potentially the first known experimental observation of the QSkHE.

The present manuscript is organised as follows: in the first half of the manuscript, we consolidate and review key examples of phenomenology of the quantum skyrmion Hall effect (QSkHE), based on results from separate works~\cite{qskhe, cook_multiplicative_2022, cook2023, cookFST2023} and three works, in particular~\cite{ay2024, winterOEPT, banerjee2024}, in preparation for discussion of the EFT in the second half of this work, Section III. Section II is fairly self-contained, however, such that readers more interested in phenomenology rather than the field theory can restrict themselves to Section II. We note, however, that the discussion of phenomenology in Section II is not exhaustive, but rather targeted towards development of the EFT specifically as a generalisation of the 2+1 D SU(2) gauge theory. 

\section{Phenomenology of the quantum skyrmion Hall effect and related states}

In preparation for discussion of the phenomenology, we first review a particularly physically intuitive mechanism of introducing non-commutativity important in understanding of the QSkHE, that of fuzzification by projection to the lowest Landau level, in Subsection A. We then consider severe fuzzification by Landau level projection in Subsection B, which is essential to understanding the nature of generalisations of quasiparticles (QPs) at the heart of generalisation from the quantum Hall effect (QHE) to the QSkHE. This motivates Subsection C, in which we discuss topological characterization of a severely-fuzzified Landau level, denoted by LL\textsubscript{F}. 

These first three sections introduce the possibility that systems with $d$ Cartesian coordinates as well as pspin DOFs, typically interpreted as realizing states that are intrinsically of dimensions $d+1$ (we denote space + time dimensions by $d+1$ and the manifold by $M^{1,d}$) or less, instead can also realize states most similar to those expected for systems with $d+\delta$ Cartesian space coordinates, due to $\delta$ dimensions being encoded in terms of the pspin DOFs in combination with $d$ dimensions encoded by Cartesian coordinates. We therefore review the higher-dimensional topological state most relevant to understanding phenomenology of the QSkHE, which is the 4+1 D Chern insulator (CI), in Subsection D. 

In Subsection E, we use the preparation of the earlier sections to  review phenomenology of the QSkHE in related lattice tight-binding models, discussed in detail in three key separate works~\cite{winterOEPT, ay2024, banerjee2024}. We note that we represent key results contained in these separate works schematically in this section, rather than reproducing the actual results in this paper. Given this phenomenology, we discuss how the topological invariant of the 4+1 D CI, the second Chern number, could emerge in lower-dimensional systems as ostensibly lower-dimensional topological invariants computed from pspin expectation values, in Subsection F. We then end Section II with a brief summary in Subsection G.

\subsection{Fuzzification by projection to the lowest Landau level}

Here, we review how non-commutativity follows directly from Landau level (LL) projection, which provides physical motivation for the generalized dimensional reduction procedure used to derive the EFT for the QSkHE.
Electrons moving on a sphere in the presence of a U(1) monopole background, form LLs and can be described by the Hamiltonian \cite{Haldane:1983xm, Hasebe:2010vp,Zhu:2022gjc},
\begin{equation}
    H=-\frac{1}{2m}D_iD^i=\frac{1}{2mr^2}\Lambda_i^2
\end{equation}
where $\Lambda_i=-i\epsilon^{ijk}x_jD_k$ and $D_i=\partial_i+iA_i$ is the covariant derivative with gauge field $A_i$ encoding effects of the monopole. 


The eigenstates of the system are described by the modified angular momentum algebra,
\begin{equation}
    [L_i,L_j]=i\epsilon_{ijk}L_k,\qquad L_i=\Lambda_i+\frac{s}{r}x_i,
\end{equation}
and the energy eigenvalues $E=\frac{1}{2mr^2}[n^2+n(2s+1)+s]$ are $(2n+2s+1)$ fold degenerate. Here, $n$ is the LL index and $2s$ is the strength of the monopole. In the Lowest Landau Level (LLL), i.e, $n=0$, the degenerate states are given by the monopole harmonics \cite{Wu:1976ge},
\begin{equation}
    \braket{\theta,\phi|s,m}=\mathcal{N}_me^{im\phi}\cos^{s+m}\left(\frac{\theta}{2}\right)\sin^{s-m}\left(\frac{\theta}{2}\right)
\end{equation}
where, $\mathcal{N}_m=\sqrt{\frac{(2s+1)!}{4\pi (s+m)!(s-m)!}}$, with $m=-s,\dots, s$. We can see the states are $2s+1$-fold degenerate. 

In the LLL, the kinetic energy is quenched with respect to the interaction term and, projecting the states onto the LLL, one can replace the position co-ordinates by the angular momentum operators, or $x_i \cong \frac{r}{s}L_i$, such that the non-commutativity of the co-ordinates is encoded in the $SU(2)$ algebra with irreducible representations labelled by $s$, 
\begin{equation}
    [x_i,x_j]=i\epsilon_{ijk}\frac{r}{s}x_k
\end{equation}

This non-commutativity via projection into the LLL defines a fuzzy sphere regularisation of $S^2$~\cite{Zhu:2022gjc} and therefore is one avenue to dimensional reduction by fuzzification. Rather than an indefinite sum over spherical harmonics and continuous co-ordinates, the non-commutativity of the fuzzy sphere~\cite{madore1992, balachandran2006} and monopole harmonics can be a used as a well-defined cutoff and a regularisation scheme.


We define co-ordinates over fuzzified spatial dimensions by embedding them in the representation of the $SU(2)$ algebra, which is labelled by the monopole strength $s$ and the basis monopole harmonics. We therefore represent the projected LLL co-ordinates in the given basis as,
\begin{equation}
    x^i=\sum_{m,m'}\bra{s,m}x^i\ket{s,m'}\ket{s,m}\bra{s,m}
\end{equation}

One can show that the radius of the fuzzy sphere, $r_F$, is defined through the Casimir or the magnitude of total angular momentum operator~\cite{madore1992, balachandran2006, Hasebe:2010vp},
\begin{equation}
   r_F= \sqrt{x_i^2}=r\sqrt{1+\frac{1}{s}}
\end{equation}
so, upon taking $s\to\infty$, we recover the two-sphere $S^2$. For the thermodynamic limit to be meaningful, the magnetic length,
\begin{equation}
    l^2=r^2/s
\end{equation}
should be finite\cite{Hasebe:2010vp, zhanghu2001}. One can see, for a nearly $0$ D case approaching the limit $r\to 0$, the finite energy gap or the magnetic length demands severe fuzziness of the sphere approaching the limit $s\to 0$. One can check that, even on vanishing physical length, the radius of the fuzzy sphere goes to a finite value, $r_F\to l$.\\

While fuzzification by Landau level projection is an intuitive and physically well-motivated approach in the present context, other variations on fuzzification are also discussed in the literature~\cite{balachandran2006}, so we discuss them here briefly, and later expand this discussion in the field theory section, Section III. As we will re-examine the effective field theories of the 4+1 D and 6+1 D QHE, we note that fuzzification is also performed more directly on NL$\sigma$M fields, and specifically in a case where NL$\sigma$M field expectation values are fixed and trivial~\cite{bernevig6Dfieldtheory}. In this work, a $U(1)$ field theory over the complex projective space $CP^3$, or effectively the six-sphere $S^6$, is reduced to an $SU(2)$ non-Abelian field theory over the four-sphere $S^4$ by taking advantage of $CP^3$ being locally isomorphic to $S^2\times S^4$. NL$\sigma$M fields over $S^2$ are fuzzified to eliminate two Cartesian coordinates. 

\subsection{Landau level on severely-fuzzified sphere}

Defining weak fuzzification as $s$ large, as considered in past work~\cite{Zhu:2022gjc}, and negligible fuzzification as the limit of $s \rightarrow \infty$ as in the previous section, we now consider the physical meaning of severe fuzzification, corresponding to small $s$ and taking $2s=1$, $2$, or $3$, in particular, as the most severe instances of fuzzification considered here. Here, we ask specifically, if a LL forms for $s$ large but finite, what is the fate of the LL upon reducing $s$ to be this small yet finite (and changing radius $r$ as required to maintain a fixed ratio of particle number to flux quanta number), corresponding to severe fuzzification?
 

Some insight into the physical meaning of a LL subjected to severe fuzzification emerges from the recently-discovered QSkHE and three sets of related topological phases of matter studied in lattice tight-binding models~\cite{qskhe}: these phases of matter are consistent with the possibility that, in general, the LL remains meaningful even under severe fuzzification, in the sense that it can still host an intrinsically 2+1 D topologically non-trivial state. 

In the following sections, we will argue that this scenario is consistent with the existence of the finite-size topological phases~\cite{cookFST2023}, which are intrinsically $\Delta$-dimensional ($\Delta$ D) topological phases persisting even when a system is only thermodynamically-large in $d<\Delta$ directions. System size is comparable to the characteristic decay length scale of boundary modes in the remaining $\Delta- d$ directions. These states are distinguished by intrinsically $\Delta$ D topological response signatures co-existing with $\Delta$-$d$ D gapless boundary modes naively associated with an intrinsically $d$ D topological state, but are distinguished from such boundary modes by effects of external fields and details of localisation.

The compactified many-body state can furthermore play the role that charged particles play in the QHE, the key generalisation of the QSkHE~\cite{qskhe}. We will argue that this notion of compactified many-body topological state generalising the notion of a charged particle is consistent with existence of the topological skyrmion phases of matter~\cite{cook2023, qskhe}, which are incompressible states for particular pspin angular momentum DOFs in systems also possessing other pspin DOFs. 

We can then furthermore potentially associate a species of compactified many-body state with each pspin DOF of a system. We will argue that this is consistent with the multiplicative topological phases (MTPs)~\cite{cook_multiplicative_2022}, which are realised for Hamiltonians with symmetry-protected tensor product Hilbert spaces similar to the Fock space of many particles. One can then realise an incompressible state for each individual pspin DOF. Each pspin incompressible state corresponds to each pspin DOF encoding an effective single-particle system, with the particles encoded in terms of that pspin angular momentum. The full system then encodes correlations between these pspin systems.

\subsection{Topological invariant on severely fuzzified sphere}

Past work has explored the fate of the topological invariant for cases of weak fuzzification: the Chern number, for instance, has been considered in non-commutative field theories realised by fuzzification, in which commuting Cartesian coordinates are replaced by non-commuting fuzzy coset space coordinates~\cite{aschieri2007}. The Chern number was found to quantize only for the non-fuzzy limit~\cite{balachandran2006}, and deviate from quantized values even for weak fuzzification. This naively suggests the invariant is unquantized and possibly ill-defined in the regime of severe fuzzification and small $s$. Given evidence indicating that intrinsically $d$+$\delta$+1 D topological states remain meaningful even with compactification of $\delta$ dimensions, however~\cite{qskhe, cookFST2023}, we conjecture a preliminary alternative definition of the topological invariant for the LL in this section more suitable for their characterization under severe fuzzification, as the central extension of the momentum operator commutator. We then return to this issue in Section III and make our definition more precise.


We therefore review derivation of the central extension of the momentum commutator for the case of formation of a 2+1 D LL defined for a  NL$\sigma$M. We employ the analogy between the Landau problem and spin precession~\cite{hasebe2024} to study the problem from the perspective of the quantum ferromagnet. This physical picture associated with the Landau problem is useful, as it provides deeper insight into the physical meaning of the central extension in the case of severe fuzzification: we will find that the central extension of the momentum operator commutator encodes the topological charge of quantum skyrmions, in the sense that they are defined in terms of a smooth winding of NL$\sigma$M fields before fuzzification, and defined in terms of position and momentum operator commutators in the case of severe fuzzification. We can identify these quantum skyrmions with LL\textsubscript{F}s.

As shown in ref.~\cite{HitoshiCentralExt}, the ferromagnetic skyrmions defined through the map of continuous spin configurations in 2+1 D, $\vec{n}(\vec{u},t):(\vec{u},t)\to S^2$, can be described by the Lagrangian
\begin{equation}
    \mathcal{L}=I\rho\Dot{\phi}(\cos{\theta}-1)-\frac{J}{2}\nabla_i n\nabla^i n.
\end{equation}
Here, $(\theta,\phi)$ parameterise the sphere $\vec{n}^2=1$ and $I\rho$ is the magnetization density. The spin configurations can be characterized by the winding number $\mathcal{Q}$ through non-trivial homotopy group $\pi_2(S^2) = \mathbb{Z}$. 

Taking $p_{\phi}=\partial L/\partial\Dot{\phi}=I\rho (\cos\theta-1)$, with
\begin{align}
    \left[ \phi(\vec{u},t),p_{\phi}(\vec{u}',t)\right]=i\delta^2(\vec{u}-\vec{u}'),
\end{align}
we use translational invariance of the action to define the momentum density $T^{0i}=p_\phi\partial_i \phi$ and corresponding momentum operator $P^i=\int \dd^2 u T^{0i}$. 

Using the relation $ \left[P^x,\phi(\vec{u},t)\right]=i\partial_x \phi,\ \left[P^x,p_\phi(\vec{u},t)\right]=i\partial_x p_\phi$ coming from  $P^i$ being the generator of translations, we obtain:
\begin{align}
    &\left[P^x,T^{0y}(\vec{u},t)\right]= \left[P^x,p_\phi(\vec{u},t)\partial_y\phi(\vec{u},t)\right]\\
    &=\left[P^x,p_\phi(\vec{u},t)\right]\partial_y\phi(\vec{u},t)+p_\phi(\vec{u},t)\left[P^x,\partial_y\phi(\vec{u},t)\right]\\
    &=i\partial_x p_\phi\partial_y\phi(\vec{u},t)+ip_\phi(\vec{u},t)\partial_x\partial_y\phi(\vec{u},t)\\
    &=i\partial_x(p_\phi\partial_y\phi(\vec{u},t))+ip_\phi(\vec{u},t)\partial_{[x}\partial_{y]}\phi(\vec{u},t),
\end{align}
so,
\begin{equation}
     \left[P^x,T^{0y}(\vec{u},t)\right]= i\partial_xT^{0y}+ip_\phi \epsilon_{ij}\partial_i\partial_j\phi(\vec{u},t).
\end{equation}
Here, we used the fact that $\phi$ has a vortex singularity coming from the $U(1)$ charge inside the sphere, which translates to a vortex singularity of the skyrmion. A consequence of this is that the derivatives do not always commute. Integrating over the second momentum density, one obtains:
\begin{align}
      \left[P^x,P^y\right]&=-\int \dd ^2 u \ i\partial_ip_\phi \epsilon_{ij}\partial_j\phi(\vec{u},t)\\
      &=iI \rho \int \dd ^2 u \  \epsilon_{ij}\sin \theta \partial_i\theta \partial_j\phi\\
    \left[P^x,P^y\right]  &=4\pi i I \rho \mathcal{Q}
      \end{align}
where,
\begin{equation}
    \mathcal{Q}=\dfrac{1}{4\pi} \int \dd^2 u \ \vec{n}\cdot \partial_x \vec{n}\times \partial_y \vec{n} \  \in \mathbb{Z}.
\end{equation}
Similar terms arise when considering the Lagrangian for a particle in the LLL over $S^2$ via the action
\begin{equation}
    S=\int dt A_a d\Dot{x}^a
\end{equation}
for $x_a^2=1$ and $A_a\sim \frac{-I}{2(1+x_3)}\epsilon_{ab3}x^b$. We use the parametrization, 
\begin{align}
    x_1&=\sin{\theta}\cos{\phi}\\
    x_2&=\sin{\theta}\sin{\phi}\\
    x_3&=\cos{\theta}
\end{align}
to obtain the Berry phase term as in the quantum ferromagnet case as,
\begin{equation}
    \mathcal{L}= I\Dot{\phi}(\cos{\theta}-1),
\end{equation}
corresponding to a central extension of the momentum commutator in this case, also.

In this example, we see we can identify skyrmions forming in the texture of the NL$\sigma$M fields with projection to a LL in the Landau model. This suggests a physical picture of the many-body state associated with filling some orbitals of the LL\textsubscript{F} as a magnetic skyrmion. 

This relation between a LL and a skyrmion is actually quite generic, as can be seen from more general relation of these two types of problems. For instance, one can promote the LLL co-ordinates in terms of NL$\sigma$M fields~\cite{Palumbo:2021dkx} $x^a=x^a(\vec{u},t)$, by defining an appropriate density $\rho(u)$.  
Using the Hopf map~\cite{Hasebe:2010vp, Hopf1931} $\vec{x}=\psi^{\dagger}(\vec{u},t)\vec{\sigma}\psi(\vec{u},t)$, where $\psi=\{\psi_1,\psi_2\}$ is the Hopf spinor with normalisation condition $\psi^{\dagger}\psi=1$ and $\vec{\sigma}$ is the vector of Pauli matrices, the action can be written as,
\begin{equation}
S=\int dt d^2u \rho(u)A_a d\Dot{x}^a=\int dt d^2u \rho(u)\psi^{\dagger}d\Dot{\psi},
\end{equation}
where we assume constant density $\rho(u)=\rho$ for simplicity, and set it to unity to simplify discussion. 

One can check the Berry curvature can be written as\cite{Palumbo:2021dkx,Wu:1988py},
\begin{equation}
  F=dA=Tr(xdx\wedge dx),
\end{equation}
where $x=x^a\sigma_a$ and the trace is with respect to the Pauli matrices. A current of point-like skyrmions, $J_{Sk}$, can similarly be expressed in terms of $F$ as 
\begin{equation}
    J_{Sk}=\star F,
\end{equation}
where $\star$ is the Hodge dual. The topological charge of the skyrmion is then
\begin{equation}
    Q=\int_{\vec{u}} \star J_{Sk}=\int_{S^2}F
\end{equation}
with proper normalisation. We note that similar relations can be constructed using higher Hopf maps\cite{Wu:1988py}.\\

\begin{figure}[t]
\includegraphics[width=0.45\textwidth]{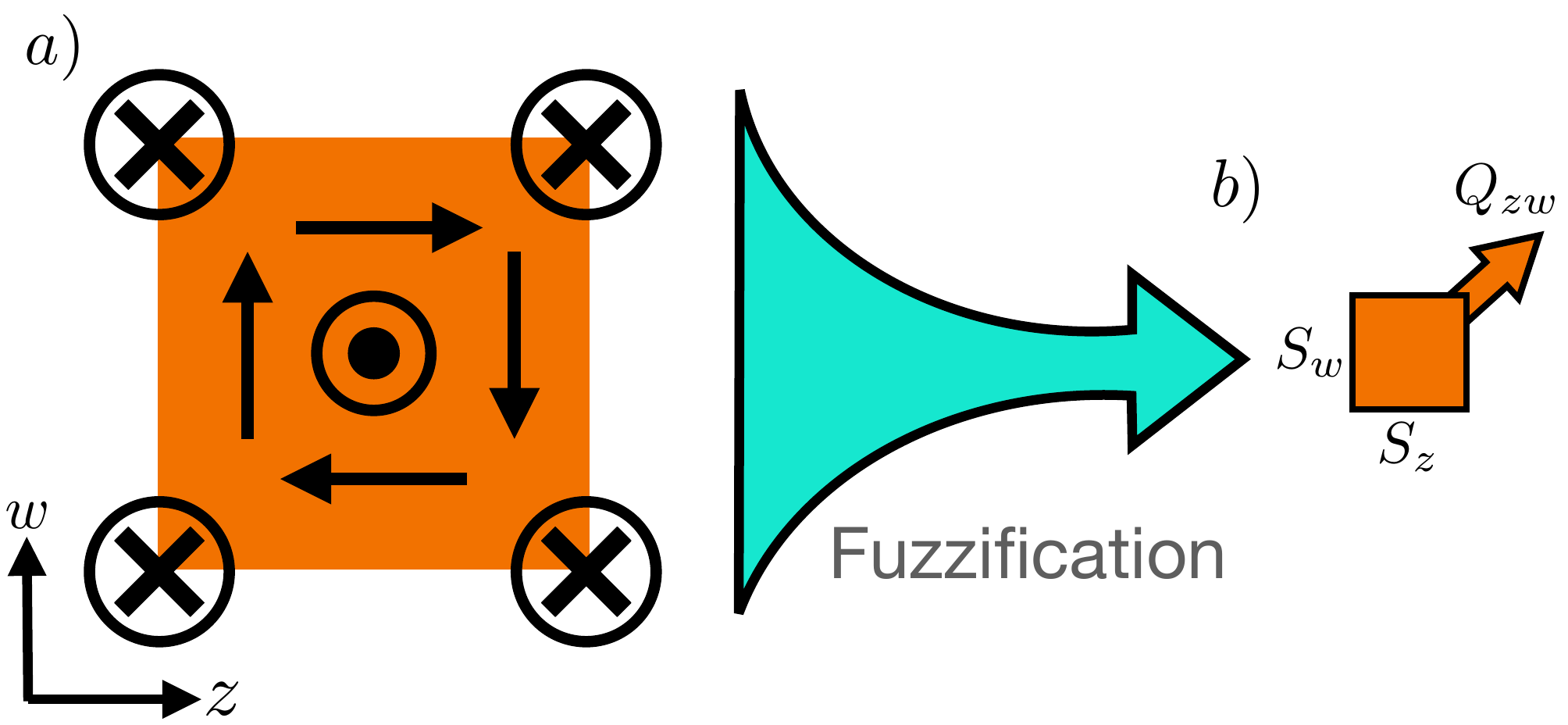}
\caption{Schematic representation of a) Landau level, LL, as smooth topological texture in NL$\sigma$M fields, and b) severely-fuzzified Landau level, LL\textsubscript{F}, with position coordinates encoded in SU(2) generators $S_z, S_{\omega}$, with topological charge of the skyrmion forming over the NL$\sigma$M fields, $\mathcal{Q}_{zw}$, encoded in the central extension of the momentum commutator.}
\label{figLLandLLf}
\end{figure}

That the topological charge of the LL is encoded in the momentum commutator provides insight into how a LL\textsubscript{F} may still effectively encode a state more akin to an intrinsically 2+1 D topological state than a 0+1 D topological state in agreement with evidence for the QSkHE. In correspondence with the notion of a LL remaining meaningful in a sense, by still characterising an intrinsically 2+1 D topologically non-trivial many-body state on the severely-fuzzified sphere (see Fig.~\ref{figLLandLLf}), gauge fields defined over severely-fuzzified two-spheres more generally also can encode 2+1 D states. Intrinsically $d + \delta$-spatial dimensional terms in quantum field theories then remain important to understanding of the physics of a system with $d$ Cartesian coordinates and $\delta$ severely-fuzzified dimensions encoded in isospin DOFs. This is the basis of the phenomenological EFT of the QSkHE introduced in this work, which is supported by evidence from lattice tight-binding models. 

\begin{figure}[t]
\includegraphics[width=0.5\textwidth]{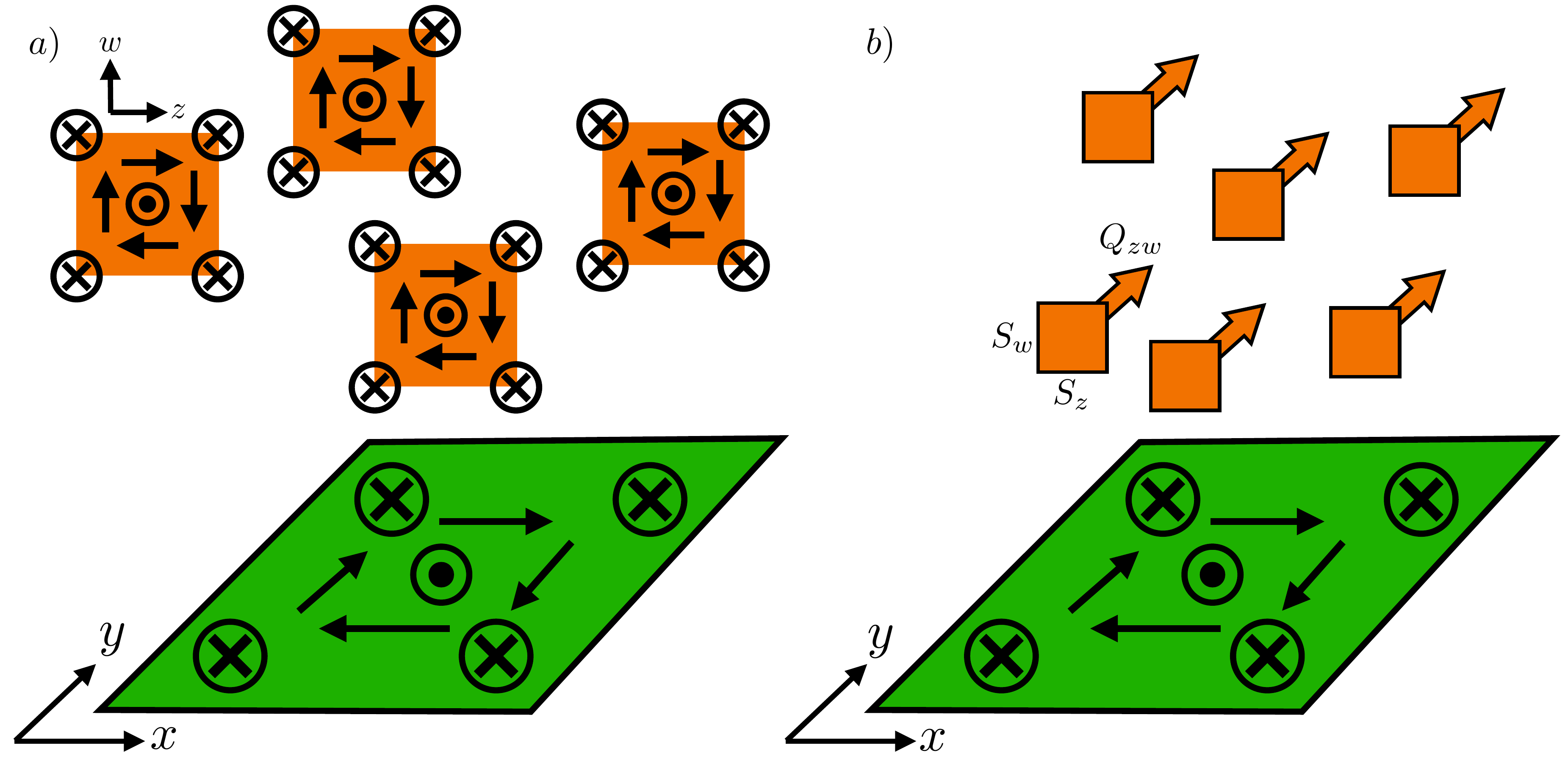}
\caption{Schematic of a) 4+1 D QHE treated heuristically as effective incompressible state of Landau levels (or equivalently membranes), shown in orange, instead of charged point particles, b) 2+1 D QSkHE treated heuristically as an effective incompressible state of compactified Landau levels.}
\label{schem_4Dqhe_2Dqskhe}
\end{figure}

This derivation of the central extension linking the LL to a magnetic skyrmion in the NL$\sigma$M also supports interpretation of the LL\textsubscript{F} as a truly quantum skyrmion, in the sense of necessarily being encoded in terms of spin angular momentum as part of retaining dependence on fuzzy dimensions, and yet carrying topological charge encoded in the non-trivial central extension of the momentum commutator. The notion of a compactified many-body state as a truly quantum skyrmion serves as the foundation of the generalised incompressible states of the QSkHE, yet is interesting in its own right, particularly with regards to non-Fermi liquid (NFL) physics. From the perspective of past study of skyrmions in terms of smooth windings of classifical fields, LL\textsubscript{F}s appear to simply be spins at first glance (see Fig.~\ref{schem_4Dqhe_2Dqskhe}), and methods must be developed to more fully characterize them as part of future work. In Section III, we reformulate the central extension derivation to identify a candidate topological invariant promising for quantisation, and employ it in characterising the most general EFT of the QSkHE presented in this work. Related to this point, we briefly note the interesting concept of a topological spin discussed in past works of Haldane and collaborators in the context of the QHE in the absence of continuous rotational symmetry \cite{haldane2016geometrylandauorbitsabsence}. It is possible that this topological spin serves as an approximation of the LL\textsubscript{F}s of the QSkHE, a topic that will be explored in future work.

\subsection{Review of the 4+1-dimensional Chern insulator}

In this subsection, we review the 4+1 D CI phase~\cite{qi2008TRIFT,bernevig2013topological}, focusing on aspects of the theory, which will be relevant to interpretation of the evidence provided by lattice tight-binding models with two Cartesian spatial coordinates for topological states that realise states best identified with known intrinsically 4+1 D---rather than intrinsically 2+1 D---topological states, motivating the EFT of the QSkHE presented in Section III.\\

A tight-binding model Hamiltonian realizing the 4+1 D CI phase can be written as~\cite{qi2008TRIFT},
\begin{equation}
    H=\sum_k \Psi^{\dagger}_kd_a(k)\Gamma^a \Psi_k,
\end{equation}
where here $\{\Gamma_a \}$ are the five Dirac matrices satisfying the Clifford algebra $\{\Gamma_a, \Gamma_b\} = 2 \delta_{ab} \mathbb{I}$, with $\mathbb{I}$ the identity matrix and the momentum $\boldsymbol{k}$ has four components, $k_x$, $k_y$, $k_z$, $k_w$, for each of the $\hat{x}$-, $\hat{y}$-, $\hat{z}$-, $\hat{w}$-directions. The $\{d_a(k)\}$ are five momentum-dependent functions treated as components of a $\boldsymbol{d}$-vector encoding specific details of a given model for the 4+1 D CI, and $\Psi_k$ is a four-component basis spinor containing second-quantized creation and annihilation operators.  

The eigenvalues of such a Hamiltonian take the form $E_{\pm}=\pm\sqrt{\sum_ad_a^2(k)}$. The topological invariant characterizing the 4+1 D CI state, the second Chern number, may be expressed as a winding number or Jacobian of the map $\hat{d}_a=\frac{d_a(k)}{\abs{d_a(k)}}:T^4\to S^4$, where $T^4$ is the four-torus and $S^4$ is the four-sphere, as
\begin{align}
    C_2 = \frac{3}{8\pi^2}\int d^4k \varepsilon^{abcjk}\hat{d}_a \partial_x \hat{d}_b \partial_y \hat{d}_c \partial_z \hat{d}_j \partial_w \hat{d}_k
\end{align}
For open boundary conditions (OBCs) in one spatial direction, say along the $\hat{w}$-axis, one expects $\abs{C_2}$ branches of chiral gapless boundary modes in correspondence with the non-trivial topology of the bulk characterized by the second Chern number. 

A point very important to later discussion of numerical results, is that $C_2$ may be non-trivial even in the presence of time-reversal symmetry (TRS), unlike the first Chern number, yet TRS does not protect the topological state, including gaplessness of boundary modes, unlike the case of the quantum spin Hall insulator (QSHI) and strong topological insulator (STI)~\cite{qi2008TRIFT}. The gapless boundary modes are 3D Weyl fermions, and can be described by the low-energy, effective Hamiltonian as~\cite{Creutz:2000bs},
\begin{equation}
    H=\text{Sgn}(C_2)\int \frac{d^3k}{(2\pi)^3}\sum_{i=i}^{\abs{C_2}}v_ic^{\dagger}_i(\sigma\cdot \boldsymbol{k}) c_i,
\end{equation}
where $\boldsymbol{k}$ is the momentum in the directions with periodic boundary conditions (PBCs) and Sgn($C_2$) defines the chirality of the edge states. One can couple this boundary theory with a $U(1)$ background field to realise the chiral anomaly response signature of the 3D Weyl nodes~\cite{qi2008TRIFT}. 

The topological response signature of the 4+1 D CI more generally is particularly relevant to later discussion and motivates very minimal discussion of the relevant Chern-Simons (CS) theory in this section focused on phenomenology. The action for the response theory of the 4+1 D CI takes the form
\begin{equation}
    \mathcal{S}=\frac{C_2}{24\pi^2}\int d^4x dt \epsilon^{\mu\nu\rho\sigma\tau}A_{\mu}\partial_{\nu}A_{\rho}\partial_{\sigma}A_{\tau},
\end{equation}
where the coefficient $C_2$ is the second Chern number, which can also be expressed in terms of non-Abelian Berry curvature over the 4D Brillouin zone (BZ) as,
\begin{equation}
   C_2=\frac{1}{32\pi^2}\int d^4k\epsilon^{ijkl}Tr[f_{ij}f_{kl}] 
\end{equation}
with the field strength $f_{ij}$ given in terms of the gauge field $a$ by,
\begin{equation}
    f^{\alpha\beta}=da^{\alpha\beta}+i (a\wedge a)^{\alpha\beta},\quad a^{\alpha\beta}=-i\bra{\alpha,k}d\ket{\beta,k},
\end{equation}
where $\alpha,\beta$ indicate occupied bands. The second Chern number $C_2$ therefore governs the response of the 4+1 D CI to external electric and magnetic fields $\boldsymbol{E}$ and $\boldsymbol{B}$, respectively. From the CS action, one then finds the current density as,
\begin{equation}
    J^{\mu}=\frac{C_2}{8\pi^2}\epsilon^{\mu\nu\rho\sigma\tau}\partial_{\nu}A_{\rho}\partial_{\sigma}A_{\tau},
\end{equation}
which is proportional to a product of electric and magnetic fields, implying  invariance of the equation under time reversal. Furthermore, while the gapless boundary modes of the QSHI and STI can gap out immediately upon introducing a perturbation breaking time-reversal symmetry, the boundary modes associated with non-trivial second Chern number $C_2$ are more robust, being 3D Weyl nodes~\cite{qi2008TRIFT}. 

We assume the simplified field configuration with $F_{zt}=-E_z$ and $F_{xy}=B_z$, where $E_z$ ($B_z$) is an electric (magnetic) field of strength $E_z$ ($B_z$) applied in the $\hat{z}$-direction. This results in the current,
\begin{equation}
    J_{v}=\frac{C_2}{4\pi^2}B_z E_z.
\end{equation}
The presence of magnetic field through the $x$-$y$ plane results in formation of degenerate LLs over the $x$-$y$ plane. On integrating out the $x$ and $y$ dependence, we have
\begin{equation}
    j_v=\int dx dyJ_{\nu}=\frac{C_2}{2\pi}N_{xy}E_z,
\end{equation}
where $N_{xy}$ is the number of flux quanta $\phi_0$ through the $x$-$y$ plane. One can see the equation implies a quantised Hall conductance $\sigma_{zw}$ in the $z$-$w$ plane, corresponding to entire LLs defined over the $x$-$y$ plane being transported in the $z$-$w$ plane in response to application of the external fields. Importantly for later discussion, then, we can interpret LLs as generalizations of particles in understanding this 4+1 D topological state. Related to this, while the boundary modes of the 4+1 D CI are gapless and chiral Weyl nodes described by effective 3+1 D theories for the boundaries, the boundary modes can also be interpreted as 2+1 D LLs propagating in a third direction, another sense in which a LL can be interpreted as a generalisation of a charged particle to understand the 4+1 D CI in relation to the 2+1 D CI and QHE.

The 4+1 D CI is a lattice counterpart of a continuum theory for the 4+1 D generalisation of the QHE  presented in\cite{Zhang:2001xs}, which is characterized by an an $SU(2)$ non-Abelian gauge theory, for particles on the four-sphere $S^4$. In trying to keep Section II on phenomenology self-contained and more accessible, we review this field theory for the 4+1 D QHE in Section III A. We note that, while the 4+1 D CI and QHE are closely-related, there are also interesting distinctions, which actually mean the theory of 4+1 D CI is more well-behaved and physically meaningful than the theory of the 4+1 D QHE.


\subsection{Evidence for minimal phenomenological field theory of quantum skyrmion Hall effect as 4+1 D SU(2) gauge theory subjected to generalized fuzzification}

In this section, we review signatures of topological phases associated with the QSkHE. Each example is realized for an effectively non-interacting Hamiltonian with two Cartesian spatial coordinates. While such models are expected to realise topological states of intrinsically 2+1 D or less, we show that considerable insight into the bulk-boundary correspondence and response signatures of these topological states comes from interpreting them as intrinsically $d$+$\delta$+1 D topological states due to $d=2$ spatial dimensions associated with Cartesian coordinates and $\delta>0$ fuzzy spatial dimensions encoded in pspin DOFs.

We first comment on evidence provided specifically by the finite-size topological phases~\cite{cookFST2023, calderon2023_fst}. This set of topological states, as mentioned earlier, is characterized by intrinsically $\Delta$+1 D topological response signatures co-existing with gapless boundary modes that appear to be a consequence of intrinsically $\Delta$-$d$+1 D topological states. Such topological phases have already been observed in lattice tight-binding models for the CI~\cite{cookFST2023}, STI and QSHI~\cite{calderon2023_fst}, crystalline topological insulator~\cite{pacholski2024}, and Weyl semimetal~\cite{pal2024_fstwsm}. These states very broadly indicate that systems with $d$ Cartesian coordinates, as well as some layer pspin degree(s) of freedom associated with Lie algebra generators of matrix representation size $N \times N$, where $N$ is small (e.g., $2$, $3$, $4$, etc.) relative to characteristic decay length scales of boundary modes, realize topological states which are intrinsically $d$+$\delta$+1 D, where $\delta>0$. This is consistent with intrinsically higher-dimensional topological states persisting after compactification by converting some Cartesian coordinates to pspin coordinates, such as by projection to the LLL. This motivates understanding of systems with such pspin DOFs ($N$ small) in terms of compactified higher-dimensional gauge theories of Section III. In this section, we instead focus specifically on evidence from topological skyrmion phases and MTPs in support of a minimal gauge theory for the 2+1 D QSkHE being the compactified 4+1 D SU(2) gauge theory.

We first introduce the form of Hamiltonians considered in this section. Throughout this section, we will consider topological states realised for four-band Bloch Hamiltonians for effectively non-interacting systems. The Hamiltonians are therefore of the general form

\begin{align}
    \mathcal{H} = \sum_{\boldsymbol{k}} \psi^{\dagger}_{\boldsymbol{k}} H(\boldsymbol{k}) \psi^{}_{\boldsymbol{k}},
\end{align}
where $H(\boldsymbol{k}) $ is the Bloch Hamiltonian and $\boldsymbol{k} = \left(k_x, k_y\right)$ is defined for the case of two Cartesian coordinates. The four-component basis spinor $\psi^{}_{\boldsymbol{k}}$ takes the form
\begin{align}
    \psi^{}_{\boldsymbol{k}} = \left( c_{\boldsymbol{k},\alpha,\beta}, c_{\boldsymbol{k},\alpha',\beta}, c_{\boldsymbol{k},\alpha,\beta'}, c_{\boldsymbol{k},\alpha',\beta'}\right)^{\top},
\end{align}
where $c_{\boldsymbol{k},\alpha,\beta}$ annihilates a fermion with momentum $\boldsymbol{k}$, spin angular momentum $\alpha$ and orbital angular momentum $\beta$, where $\{\alpha, \alpha' \}$ and $\{\beta, \beta' \}$ are two two-fold pspin DOFs. In the case of each example, we will further specify the Bloch Hamiltonian and basis spinor.

We now review key concepts of topological skyrmion phases of matter and the QSkHE in preparation for considering the examples of phenomenology in the lattice tight-binding models relevant specifically to the SU(2) gauge theory of the QSkHE. The first known topological
skyrmion phases are $2+1$ D topological states possible in
effectively non-interacting systems~\cite{cook2023}, which are
characterized by the topological invariant $\mc{Q}$, the topological charge of
winding in the texture of a pspin expectation value of the occupied states over the BZ, defined as
\begin{equation}
    \mc{Q} = \frac{1} {4\pi} \int_{BZ}d\boldsymbol{k} \left[ \langle \boldsymbol{\hat{S}}(\boldsymbol{k}) \rangle \cdot \left(\partial_{k_x}\langle \boldsymbol{\hat{S}}(\boldsymbol{k}) \rangle \times \partial_{k_y}\langle \boldsymbol{\hat{S}}(\boldsymbol{k}) \rangle \right) \right],
    \label{skyrmnum}
\end{equation}
\\
where $\boldsymbol{S} = \left(S_1, S_2, S_3\right)$ is the spin representation
and $\langle \boldsymbol{\hat{S}}(\boldsymbol{k}) \rangle$ is the normalized
expectation value of the spin for occupied states. Here $\langle
S_i(\boldsymbol{k}) \rangle= \sum_{n \in \mathrm{occ}} \langle n,
\boldsymbol{k} | S_i | n, \boldsymbol{k} \rangle $, with $i \in \{1,2,3\}$ and
$| n, \boldsymbol{k} \rangle$ the Bloch state associated with the
$n$\textsuperscript{th} band. For systems with a single pspin DOF, $\mc{Q}$ is locked in value to the value of the total Chern number, but decouples from the total Chern number in systems with
multiple DOFs~\cite{cook2023}.

We first discuss the bulk-boundary correspondence of topological skyrmion phases and MTPs of matter. We start with the Bernevig-Hughes-Zhang (BHZ) model~\cite{QSHI-HgTe-Theory} and later review the $\mathcal{C}'$ four-band model for topological skyrmion phases~\cite{winterOEPT}, and  the four-band model for the multiplicative Chern insulator (MCI)~\cite{banerjee2024}. We will argue that these signatures support interpretation of the skyrmion number $\mathcal{Q}$ as a compactified second Chern number, $C_2$, the topological invariant of the 4+1 D CI, a point explored in greater detail in Section III.

\subsubsection{Bulk-boundary correspondence of the BHZ model subjected to weak applied Zeeman field}

Here, we summarize results on topological skyrmion phases of matter of the Bernevig-Hughes-Zhang model Hamiltonian for HgTe quantum wells \cite{QSHI-HgTe-Theory}, discussed in detail in separate work Ay~\emph{et al.}. The Bloch Hamiltonian $H(\boldsymbol{k})$ is taken to be $H_{BHZ}(\boldsymbol{k})$, which has the form \cite{QSHI-ham}:
\begin{align}
    H_{BHZ}(\boldsymbol{k}) &= d_z(\boldsymbol{k}) \tau^0 \sigma^z + d_x(\boldsymbol{k}) \tau^z \sigma^x + d_y(\boldsymbol{k}) \tau^0 \sigma^y \\ \nonumber
    &+ c_A \tau^x \sigma^y + c_R (f_x\left(\boldsymbol{k}\right)\tau^x \sigma^0 + f_y\left(\boldsymbol{k}\right)\tau^y \sigma^0) \\ \nonumber
    &+ \boldsymbol{h}\cdot \boldsymbol{\tau}\sigma^0.
\end{align}

Here, the $\boldsymbol{d}$-vector $\boldsymbol{d}(\boldsymbol{k}) = \left(d_x(\boldsymbol{k}), d_y(\boldsymbol{k}), d_z(\boldsymbol{k}) \right)$ is taken to have components as
\begin{align}
d_x(\boldsymbol{k}) &= \sin(k_x) \\ d_y(\boldsymbol{k}) &= \sin(k_y) \\ d_z(\boldsymbol{k}) &= u + \cos(k_x) + \cos(k_y),
\end{align}
with $u$ serving as an effective mass. $c_A$ denotes atomic spin-orbit coupling (SOC) strength. $c_R$ denotes Rashba SOC strength and $\boldsymbol{f}$ is a vector $\boldsymbol{f}(\boldsymbol{k}) = \left(f_x(\boldsymbol{k}), f_y(\boldsymbol{k}), f_z(\boldsymbol{k}) \right)$ with components $f_x(\boldsymbol{k})= \sin(k_y)$, $f_y(\boldsymbol{k})= -\sin(k_x)$ and $f_z(\boldsymbol{k})= 0$. $\boldsymbol{h}=\left(h_x, h_y, h_z \right)$ encodes an effective Zeeman field, where $\boldsymbol{\tau} = \left(\tau^x, \tau^y, \tau^z \right)$ is the vector of Pauli matrices in the $\tau$ sector. For non-negligible SOC, the established topological classification is in terms of topological invariant $\nu \in \mathbb{Z}_2$, and one may tune the system through $\nu = 0$ and $\nu \neq 0$ by tuning $u$ relative to the other free parameters of the model.

Ay~\emph{et al.} compute a skyrmion number $\mathcal{Q}$ for the BHZ model, in this case for the orbital angular momentum, using a pspin representation that combines contributions to orbital angular momentum in the two spin angular momentum sectors additively at each point in the BZ. $\mathcal{Q}$ is non-trivial and stable both in the absence of applied Zeeman field (when the system is TRI) for $\nu$ non-trivial, and under application of weak Zeeman field breaking TRS, even when the topological invariant of the QSHI, $\nu$, is ill-defined. This robustness of $\mathcal{Q}$ is very similar to that of second Chern number $C_2$ for an intrinsically 4+1 D state over two Cartesian spatial dimensions and two fuzzy spatial dimensions~\cite{ay2024}. Such an interpretation of the skyrmion number is further supported by discussion in the next section of the bulk-boundary correspondence of the BHZ model in these scenarios, upon tracing out the orbital DOF. We discuss this possible interpretation of the skyrmion number as a compactified second Chern number in greater detail in the last subsection of this first half of the manuscript focusing on phenomenology.

\begin{figure}[t]
\includegraphics[width=0.45\textwidth]{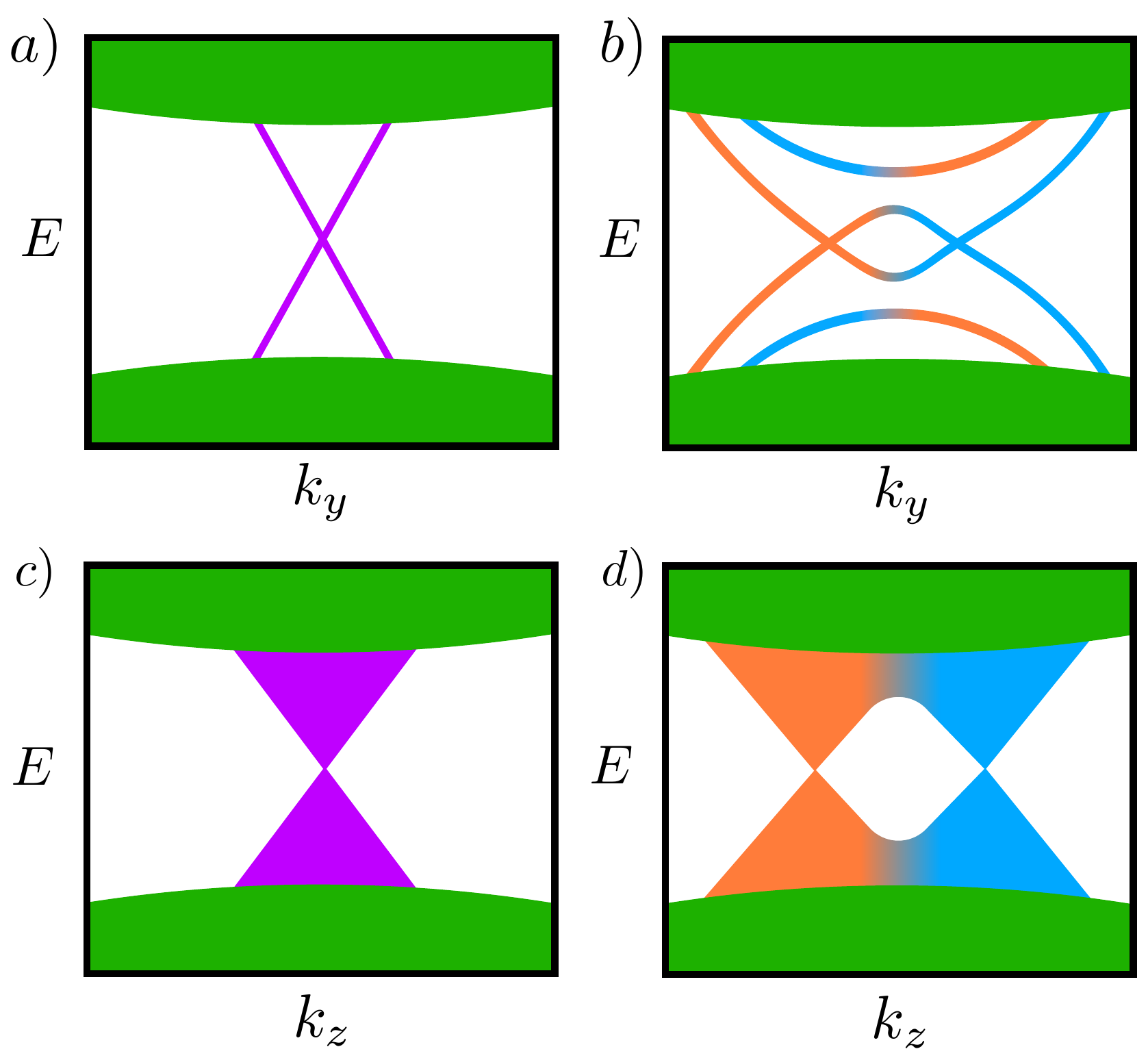}
\caption{Schematics of results from Ay~\emph{et al.}~\cite{ay2024} for a) bulk-boundary correspondence of BHZ model with TRS and b) bulk-boundary correspondence of BHZ model without TRS due to weak applied Zeeman field. Schematics of c) bulk-boundary correspondence of 4+1 D CI with TRS, and d) bulk-boundary correspondence of 4+1 D CI without TRS due to weak applied Zeeman field. Purple denotes degeneracy of states localized on each of left and right edges. Orange denotes localization of state on left edge, blue denotes localization of state on right edge, green or grey denotes bulk states. }
\label{fig_4DCI_2DBHZ}
\end{figure}

In the case of the BHZ model, one example of bulk-boundary correspondence, which serves as evidence of effective 4+1 D  topology in the ostensibly 2+1 D BHZ model, occurs for a parameter set corresponding to $\mathbb{Z}_2$ invariant $\nu$ non-trivial. Taking OBCs in the $\hat{x}$-direction for the time-reversal-invariant system yield what naively appear to be the helical edge modes of the quantum spin Hall insulator (QSHI), as shown in Fig.~\ref{fig_4DCI_2DBHZ} a). Under application of a sufficiently weak but finite Zeeman field in the $\hat{z}$-direction, normal to the x-y plane, however, the slab spectrum evolves to one similar to Fig.~\ref{fig_4DCI_2DBHZ} b). In this case, the spectrum remains gapless, but for significant subsets of the $\nu$ non-trivial regions of phase space, there is a non-negligible hybridisation gap at the center of the slab BZ ($k_y=0$) at finite energy, \textit{between bands localized on opposite edges}. Two counter-propagating bands appear within the bulk gap and localized on the left edge with one crossing point at zero energy for $k_y<0$. Near $k_y=0$, these bands \textit{delocalize} and, for $k_y >0$, they are instead localized on the right edge and crossing at zero energy for $k_y>0$. What is unexpected here is the finite hybridisation gap at $k_y=0$, realised over an extended subset of $\nu \neq0$ regions of the phase diagram as detailed in Ay~\emph{et al.}~\cite{ay2024}. 

We compare this bulk-boundary correspondence of the BHZ model under weak Zeeman field to the bulk-boundary correspondence for the 4+1 D CI for second Chern number $C_2=1$ yielding a pair of boundary Weyl nodes with charges of magnitude $1$ and opposing sign. When the system is TRS, the Weyl node localized on each boundary of the 4+1 D CI for OBCs in the $\hat{w}$ direction is located at the origin in the 3D boundary BZ, as shown in Fig.~\ref{fig_4DCI_2DBHZ} c). Breaking TRS with a weak, out-of-plane Zeeman field shifts the Weyl nodes in opposite directions in the boundary BZ as shown in Fig.~\ref{fig_4DCI_2DBHZ} d). Importantly, we note that the two boundary 3D Weyl nodes are connected by bulk states \textit{at energies less than that of the 4D bulk valence and conduction bands}, to respect the Nielsen-Ninomiya theorem \cite{Nielsen:1981hk, Friedan:1982nk} forbidding the presence of unpaired chiral fermions in the case of lattice regularisation.

We identify the two band crossing points in the bulk energy gap of the BHZ model, one on each edge, with the two boundary 3D Weyl nodes of the 4+1 D CI, which shift in the 3D boundary BZ of the 4+1 D CI  similarly under application of a weak Zeeman field, to the shift in the band-crossings observed for the BHZ model subjected to application of the weak Zeeman field. We also identify the hybridisation gap of the bulk-boundary correspondence in the BHZ model subjected to finite Zeeman field with the delocalized states connecting the boundary 3D Weyl nodes of the 4+1 D CI at energies within the bulk energy gap of the CI. The gapless points associated with crossing of edge bands in the bulk gap, for finite hybridisation gap and out-of-plane, weak Zeeman field, furthermore \textit{do not gap out} under weak rotation of the applied Zeeman field away from the $\hat{z}$-axis at arbitrary angles~\cite{ay2024}. That is, they are more robust than expected for helical edge states of the QSHI in the absence of TRS. Such robustness is actually consistent with interpretation of these band crossings as fuzzified 3D Weyl nodes, WN\textsubscript{F}s, which can also be interpreted in terms of LL\textsubscript{F}s propagating on the boundary.

Ay~\emph{et al.}~\cite{ay2024} then re-examined past experimental work on HgTe QWs~\cite{ma_unexpected_2015}, which reported unexpected edge conduction in HgTe QWs in the presence of a finite Zeeman field and also orbital magnetic field applied out-of-plane. Under these circumstances, theory of the QSHI predicts gradual loss of local density of states (LDOS) at the edge due to magnetic backscattering, up to a critical orbital field 
strength $B_c$, corresponding to loss of the bulk band inversion required to realise the QSHI state. Beyond the critical orbital field strength, no edge conduction is expected based on the theory of the QSHI~\cite{qi2008TRIFT}. In contrast to theory, however, edge conduction in the experiment persisted without gradual loss of local density of states (LDOS) at the edge, well beyond the critical orbital magnetic field strength. This effect was not explained in this past work. Modeling effects of magnetic backscattering by a finite in-plane Zeeman field component at arbitrary angle, Ay~\emph{et al.}~\cite{ay2024} examined the BHZ model for signatures of robust edge conduction in the presence of external Zeeman---and also orbital---magnetic fields, for parameter sets corresponding to finite hybrisation gap $\Delta$ and putative WN\textsubscript{F}s. Unlike helical edge modes of the QSHI, the gapless points identified as WN\textsubscript{F}s yielded finite LDOS at $E=0$ and robust edge conduction consistent with that observed in the experiments. This past work~\cite{ma_unexpected_2015} is therefore potentially the first known experimental observation of phenomenology of the QSkHE beyond the QHE.

\begin{figure}[t]
\includegraphics[width=0.4\textwidth]{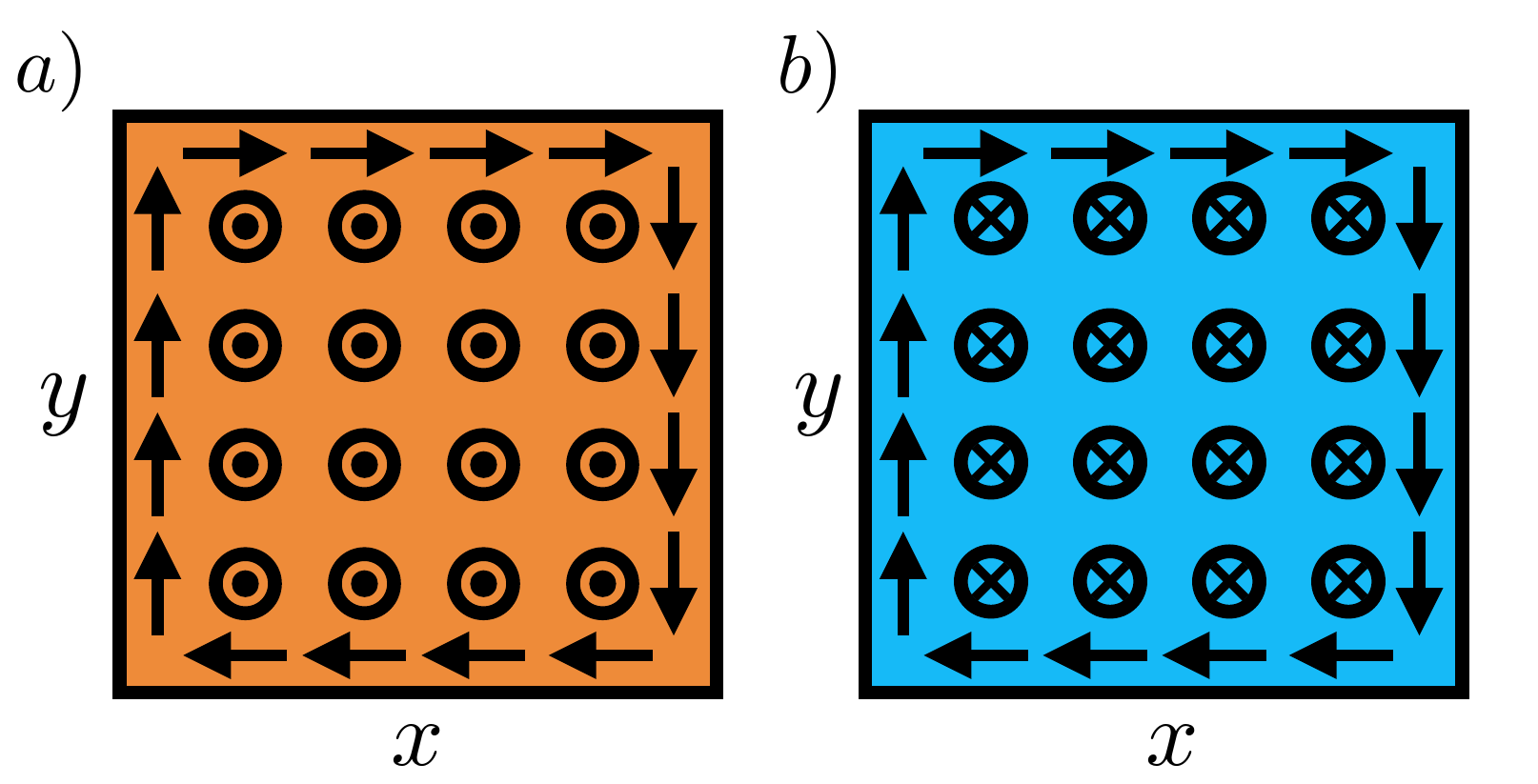}
\caption{Schematics of numerics from Ay~\emph{et al.}~\cite{ay2024} for a) real-space pspin texture for $\mc{Q}=1$ and f) real-space pspin texture for $\mc{Q}=-1$, after passing through a type-II topological phase transition starting with parameter set in a). Orange and blue, along with symbols in system bulk, indicate change in sign of out-of-plane spin component with change in sign of $\mc{Q}$.}
\label{fig_bhz_rlspin}
\end{figure}

Finally, we note that Ay~\emph{et al.}~\cite{ay2024} have also studied effects of the change in sign of the skyrmion number $\mc{Q}$ on real-space pspin textures, schematically shown in Fig.~\ref{fig_bhz_rlspin}. Under application of a weak Zeeman field, spin polarisation of occupied states is predominantly localized near---and parallel to---the boundary, with a chirality of in-plane spin components that does not change in sign with change in sign of $\mc{Q}$. However, there is a change in sign of the out-of-plane component of the spin texture upon change in sign of $\mc{Q}$. We will return to this point when discussing the example of the $\mc{C}'$ model for topological skyrmion phases of matter~\cite{winterOEPT, cook2023} 

\subsubsection{Bulk-boundary correspondence of the spin subsystem of the BHZ model}

Winter~\emph{et al.}~\cite{winterOEPT} introduces the notion of observable-enriched partial trace (OEPT), a generalization of the partial trace operation useful in further characterizing the pspin topology of the BHZ model as well as others. We briefly discuss the key results of Ay~\emph{et al.}~\cite{ay2024} employing the OEPT in analysis of the BHZ model, in particular the bulk-boundary correspondence. The OEPT $\Tilde{\mathrm{Tr}}$ with respect to an observable $S$ is an operation performed on the full density matrix of the BHZ model, $\rho$, which produces an auxiliary density matrix $\rho_s$. $\rho_s$ is constructed entirely from the expectation value $\langle S \rangle$ by satisfying the relation $\text{Tr}[\rho S ] = \text{Tr}[\rho_s \tilde{S}]$. Here, $\text{Tr}$ is the trace operation and $\tilde{S}$ is the minimal orbital angular momentum (OAM) operator matrix representation of the BHZ model in the absence of the spin half DOF, which is the vector of Pauli matrices. This defines a generalised partial trace,  $\rho_s = \Tilde{\mathrm{Tr}}[\rho]$, which preserves the expectation value of the pspin observable of interest, here OAM.  

\begin{figure}[t]
\includegraphics[width=0.4\textwidth]{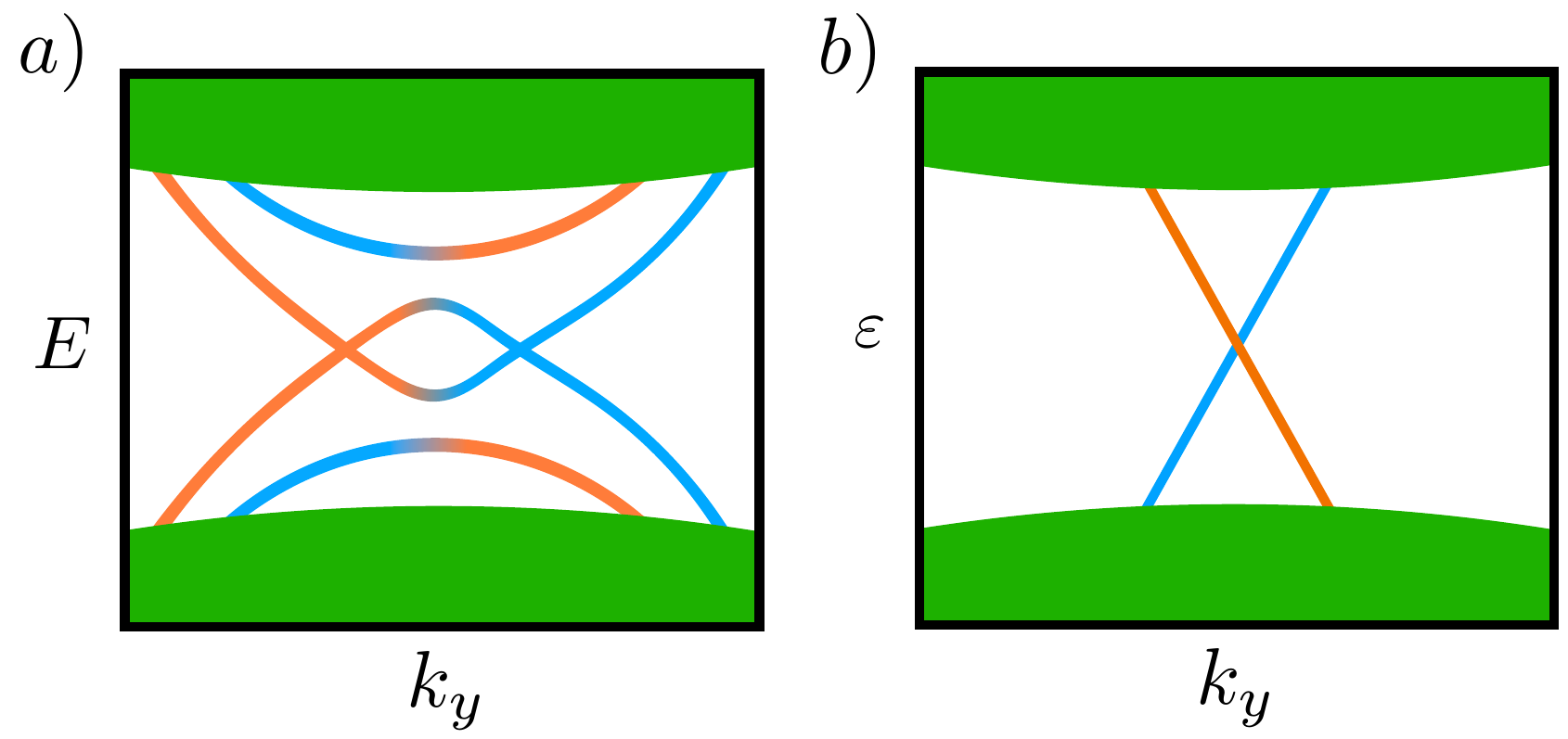}
\caption{Schematic of (a) energy slab spectrum for OBC in the $\hat{x}$-direction, PBC in the $\hat{y}$-direction for the BHZ model, with weak Zeeman field breaking TRS, (b) observable-enriched entanglement spectrum for BHZ model on a torus with additional real-space virtual cuts, for the same parameter set as in (a). Purple denotes degeneracy of states localized on each of left and right edges. Orange denotes localization of state on left edge, blue denotes localization of state on right edge, green or grey denotes bulk state localization.}
\label{fig_OEE_BHZ}
\end{figure}

The spectrum of $\rho_s$, known as the observable-enriched entanglement spectrum (OEES), is employed in Ay~\emph{et al.}~\cite{ay2024} to characterize bulk-boundary correspondence in the scenario shown in Fig.~\ref{fig_4DCI_2DBHZ} b), also shown in Fig.~\ref{fig_OEE_BHZ} a) for direct comparison with the counterpart OEES shown in Fig.~\ref{fig_OEE_BHZ} b). The OEES exhibits $\mathcal{Q}$ chiral modes localized on each entanglement cut in correspondence with the skyrmion number $\mathcal{Q}$, when combined with tracing out of half of the system in real-space for PBCs in each Cartesian real-space direction. That is, $\mathcal{Q}$ characterises a CI of the spin DOF~\cite{qskhe}.

\subsubsection{Bulk-boundary correspondence of the multiplicative Chern insulator under weak breaking of the Bloch Hamiltonian tensor product structure}

Here, we discuss recent results of related work by Banerjee~\emph{et al.}~\cite{banerjee2024} on the MCI and its implications for the EFT of the QSkHE. 

The MTPs are distinguished by symmetry-protected tensor product structure of the Bloch Hamiltonian~\cite{cook_multiplicative_2022}, meaning we consider Bloch Hamiltonians of the form $H(\boldsymbol{k}) = \mc{H}_{p1}(\boldsymbol{k}) \otimes \mc{H}_{p2}(\boldsymbol{k})$ in this example, where $\otimes$ is specifically the Kronecker product operation in this context. Non-trivial topology of $\mc{H}_{p1}$ and $\mc{H}_{p2}$ are therefore very directly inherited by $\mc{H}(\boldsymbol{k})$, which describes a phase of matter associated specifically with the tensor product structure, given that this structure is symmetry-protected. While the tensor product structure is protected by symmetries of the ten-fold way classification scheme, and therefore quite robust, the tensor product structure can be broken in a controlled manner while respecting the symmetries protecting the multiplicative structure in the bulk, such as via OBCs, leading to rich structure due to controlled entanglement, such as topologically-robust floating edge bands or gapless symmetry-protected topological phases~\cite{calderongapless}.

We consider the case of the MCI, taking $H(\boldsymbol{k}) = H_{MCI}(\boldsymbol{k}) = \mc{H}_{p1}(\boldsymbol{k}) \otimes \mc{H}_{p2}(\boldsymbol{k})$, where $\mc{H}_{p1}(\boldsymbol{k})$ and $\mc{H}_{p2}(\boldsymbol{k})$ are each two-band parent Bloch Hamiltonians for the CI state. The parent CI Bloch Hamiltonians are defined as,
\begin{equation}
\begin{split}
&\mc{H}_{p1}(\boldsymbol{k})=\mathbf{d}_1 (\boldsymbol{k})\cdot \boldsymbol{\tau};\quad \mc{H}_{p2}(\boldsymbol{k})=\mathbf{d}_2 (\boldsymbol{k}) \cdot \boldsymbol{\sigma},
\end{split}    
\end{equation}
where $\mathbf{d}_1(\boldsymbol{k})$ and $\mathbf{d}_2(\boldsymbol{k})$ are momentum-dependent, three-component vectors of scalar functions, and each of $\boldsymbol{\sigma}$ and $\boldsymbol{\tau}$ is the vector of Pauli matrices corresponding to the $\{\alpha, \alpha' \}$ and $\{\beta, \beta' \}$ DOFs, respectively. The multiplicative child Hamiltonian may more compactly be written as,
\begin{equation}
\begin{split}
& \mc{H}(\boldsymbol{k})=(d_{11},d_{21},d_{31})\cdot\boldsymbol{\tau}\otimes (-d_{12},d_{22},d_{32})\cdot \boldsymbol{\sigma},
\end{split}    
\end{equation}
to highlight the tensor product structure of the child Hamiltonian. 

Importantly, while the multiplicative Bloch Hamiltonian can be interpreted as purely quadratic in creation and annihilation operators, the Kronecker product structure also permits a second interpretation of the Hamiltonian in this case as purely quartic in creation and annihilation operators. $\mc{H}_{p1}(\boldsymbol{k})$ and $\mc{H}_{p2}(\boldsymbol{k})$ are then each interpreted as purely quadratic in creation and annihilation operators. While this dual interpretation already suggests the need to define some generalisation of a charged QP in terms of pspin DOFs, which are LL\textsubscript{F}s given previous discussion, we discuss further phenomenology of MTPs in this section which strengthens this interpretation.

We take the $\boldsymbol{d}$-vector of parent $1$, with $\boldsymbol{d} = \langle d_1, d_2, d_3\rangle$ to be that of the QWZ model
\\
\begin{align}
d_{11} &= \Delta \sin{k_y} \\ \nonumber
d_{21} &= \Delta  \sin{k_x} \\ \nonumber
d_{31} &= m-2t\left( \cos(k_x) + \cos(k_y) \right)\\ \nonumber
\end{align}
$\mathcal{H}_{p1}(\bk)$. We take $\mathcal{H}_{p2}(\bk)$ to be the time-reversed partner of $\mathcal{H}_{p1}(\bk)$, using the time-reversal operator $\mathcal{T} = i \sigma_y \mathcal{K}$, where $\sigma_y$ is the second Pauli matrix and $\mathcal{K}$ is complex conjugation, as
 \begin{equation}
 \mathcal{H}_{p2}(\bk) = \mathcal{T} \mathcal{H}_{p1}(\bk)  \mathcal{T}^{-1}. 
 \end{equation}
The tensor-product structure guarantees that the energy spectrum of the child Hamiltonian is a product of the energy spectrum of $\mc{H}_{p1}(\boldsymbol{k})$ i.e., $E_{p1}(\boldsymbol{k})$, and of $\mc{H}_{p2}(\boldsymbol{k})$ i.e., $E_{p2}(\boldsymbol{k})$, respectively,
\begin{equation}
E^c_{12}(\boldsymbol{k})= \pm E_{p1}(\boldsymbol{k})E_{p2}(\boldsymbol{k}).
\label{eq2xdegen}
\end{equation}
\\
This implies that bands of the child Hamiltonian dispersion are \textit{at least} doubly degenerate everywhere in the bulk BZ.

\begin{figure}[t]
\includegraphics[width=0.4\textwidth]{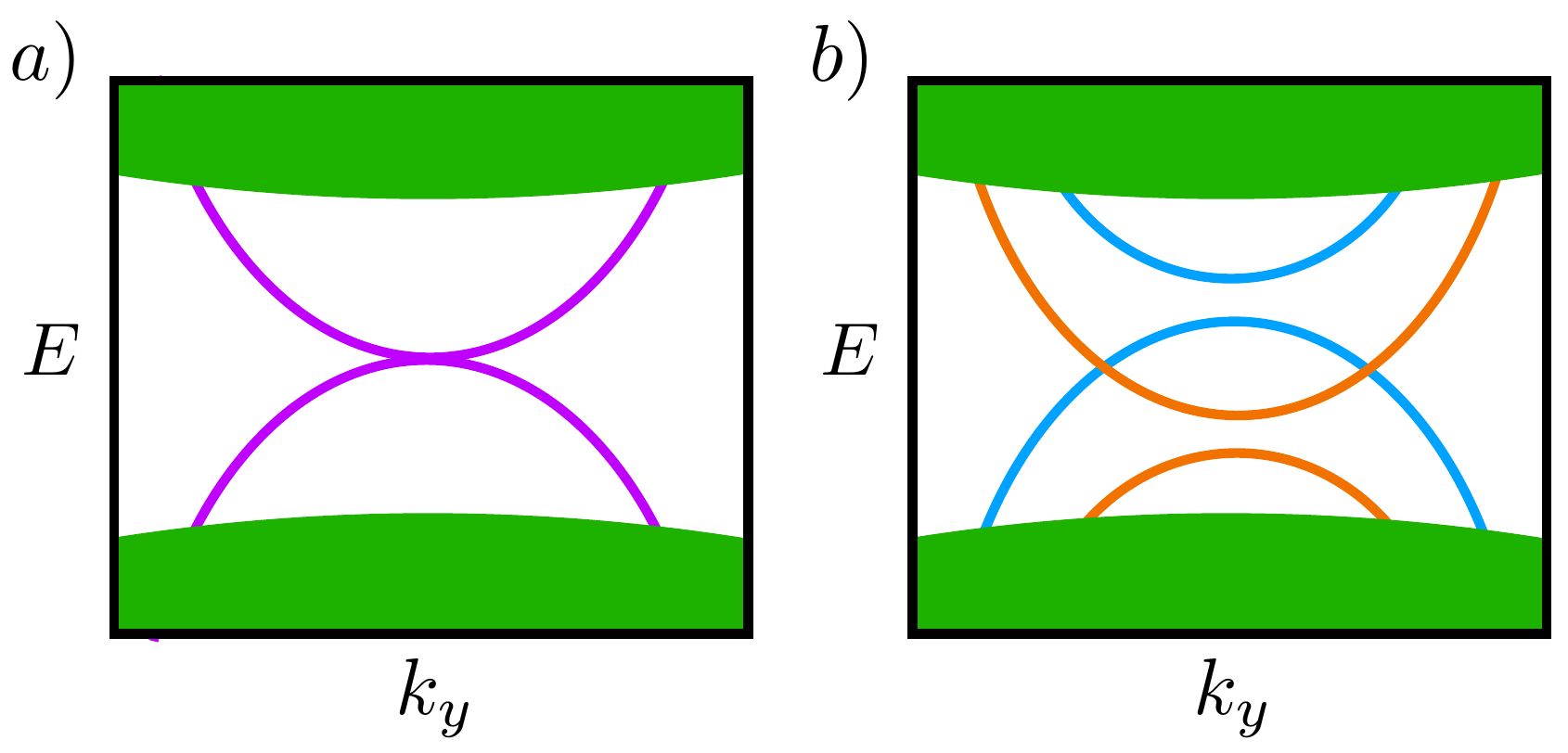}
\caption{Schematic of results from Banerjee~\emph{et al.}~\cite{banerjee2024} for the slab spectrum of the MCI a) with bulk tensor product structure, and b) with added term $H'$ breaking bulk tensor product structure. Green corresponds to bulk states, purple corresponds to degenerate bands localised on opposite edges, and blue/orange indicate localisation of states on opposite edges.}
\label{fig_MCI_slab}
\end{figure}

We focus on the case of the parent Hamiltonians each realising a CI state, with Chern numbers $+1$ and $-1$, respectively, as the parents are related by TRS.  This yields quadratically dispersing edge states in $\mc{H}(\boldsymbol{k})$ that are at least two-fold degenerate as shown in Fig.~\ref{fig_MCI_slab} a), similarly to the case of quadratically-dispersing surface modes realized for the multiplicative Hopf insulator~\cite{cook_multiplicative_2022}.  

We first discuss phenomenology of the MCI bulk-boundary correspondence, when the MCI is subjected to weak breaking of the bulk tensor product structure of $H(\boldsymbol{k})$, by addition of two terms 
\begin{align}
     \mathcal{H}'_{const} &= m' (\tau_z \sigma_z + \tau_z \sigma_0 + \tau_0 \sigma_z + \tau_x \sigma_z + \tau_y \sigma_0) \\
      \mathcal{H}'(\boldsymbol{k}) &= 2 t' \cos(k_x) \tau_y \sigma_y.
\end{align}
These terms reduce the Hamiltonian symmetry to that of class A of the ten-fold way classification scheme~\cite{schnyder2008, Ryu_2010}. Under such symmetry-breaking, the bulk-boundary correspondence evolves as shown in Fig.~\ref{fig_MCI_slab} b): the gapless points at the center of the slab BZ are gapped out, and the two-fold degeneracy of edge states is lost, save for two crossing points between bands localised on opposite edges of the system. One edge band yielding the crossings extends from the bulk valence bands and then descends back into the bulk valence bands and is localized on the left edge. The other extends from the bulk conduction bands and then descends back into the bulk conduction bands and is localized on the right edge. 

Over the interval in $k_y$ bounded by the two $k$ points at which these edge bands cross, a net change in each parent pspin expectation value is observed, predominantly for the pspin component parallel to the edge. The left edge state effectively gains approximately the amount lost by the right edge state. Similar transfer of spin angular momentum is also observed in low-symmetry three-band models for topological skyrmion phases of matter~\cite{qskhe}. 

\begin{figure}[t]
\includegraphics[width=0.35\textwidth]{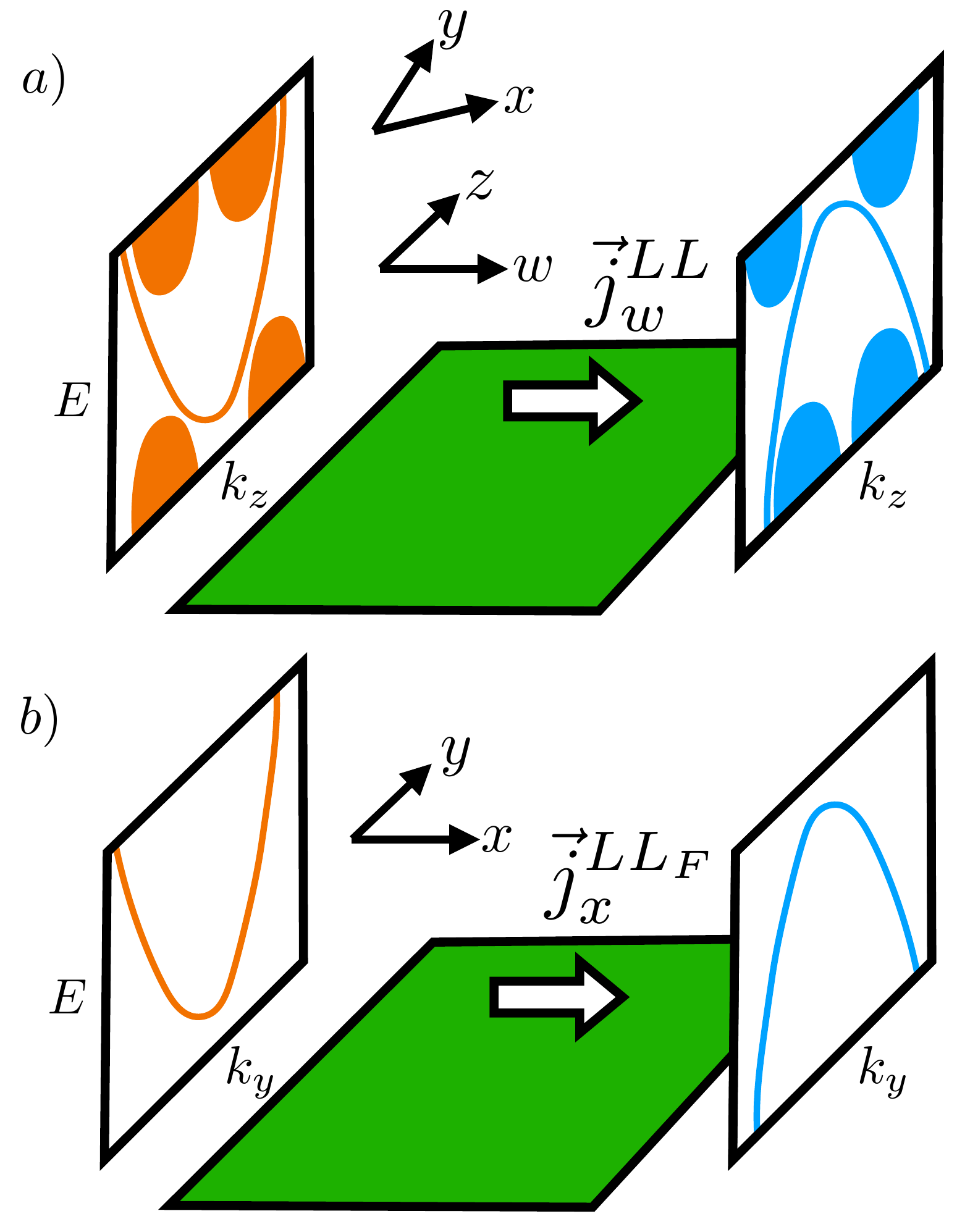}
\caption{Schematic of a) topological response signature of the 4+1 D CI with OBC in the $w$ direction. The chiral LLL is shown for a 3+1 D Weyl semimetal state localised on the left (orange) and (right) edges, respectively. $\vec{j}_w^{LL}$ denotes current density of LLs in the $w$ direction. b) Bulk-boundary correspondence of the MCI with weakly-broken tensor product structure. The in-gap bands crossing as shown in Fig.~\ref{fig_MCI_slab} b) are depicted, localised on the left (orange) and right (blue) edges, respectively. In parallel to a), $\vec{j}_x^{LL_F}$ denotes current density of LL\textsubscript{F}s in the $x$ direction.}
\label{fig_responsebklbnd}
\end{figure}

This bulk-boundary correspondence of the MCI when the tensor product structure of the Hamiltonian is broken is well understood in terms of the topological response of a 4+1 D CI with OBCs in one direction rather than known lower-dimensional topological states associated with mapping to the projector onto occupied states, or projector onto the many-body state, more generally. In the case of the 4+1 D CI, boundary 3+1 D Weyl semimetals, each with a pair of Weyl nodes of opposite chirality, form chiral Landau levels, associated with boundary chiral anomalies~\cite{qi2008TRIFT}. These boundary chiral anomalies furthermore yield a topological current density in the 4+1 D CI, in the direction of OBCs, which can be interpreted as a current density of LLs propagating across the bulk, as depicted schematically in Fig.~\ref{fig_responsebklbnd} a). In the MCI, the two crossing edge bands realised by breaking the bulk tensor product structure weakly in the bulk are identified with these boundary chiral LLs of the 4+1 D CI, and the net transfer of pspin angular momentum is identified with transport of LL\textsubscript{F}s in the direction of OBCs, in direct analogy to the transport of LLs in the case of the 4+1 D CI topological response. This bulk-boundary correspondence of the MCI---and other models realising topological skyrmion phases~\cite{qskhe}, is depicted in Fig.~\ref{fig_responsebklbnd} b).

While a topological response is normally expected upon application of external fields, the MCI and related models for topological skyrmion phases exhibit this bulk-boundary correspondence best identified with the topological response of the 4+1 D CI, in the absence of external fields. Instead, entanglement between pspin DOFs introduced by breaking the bulk tensor product structure effectively plays the role of external fields. This intertwining of bulk-boundary correspondence and topological response, also discussed in greater detail in later sections on the 6+1 D U(1) gauge theory, motivates development of potential schemes in experiment, where the amount of effective external field the system contains (leading to seeing the chiral anomaly without applying any external field) might be extracted by applying external fields to counter the effective external fields.

\subsubsection{$4 \pi$ Aharonov Bohm effect of 2+1 D multiplicative Chern insulator}

\begin{figure}[t]
\includegraphics[width=0.45\textwidth]{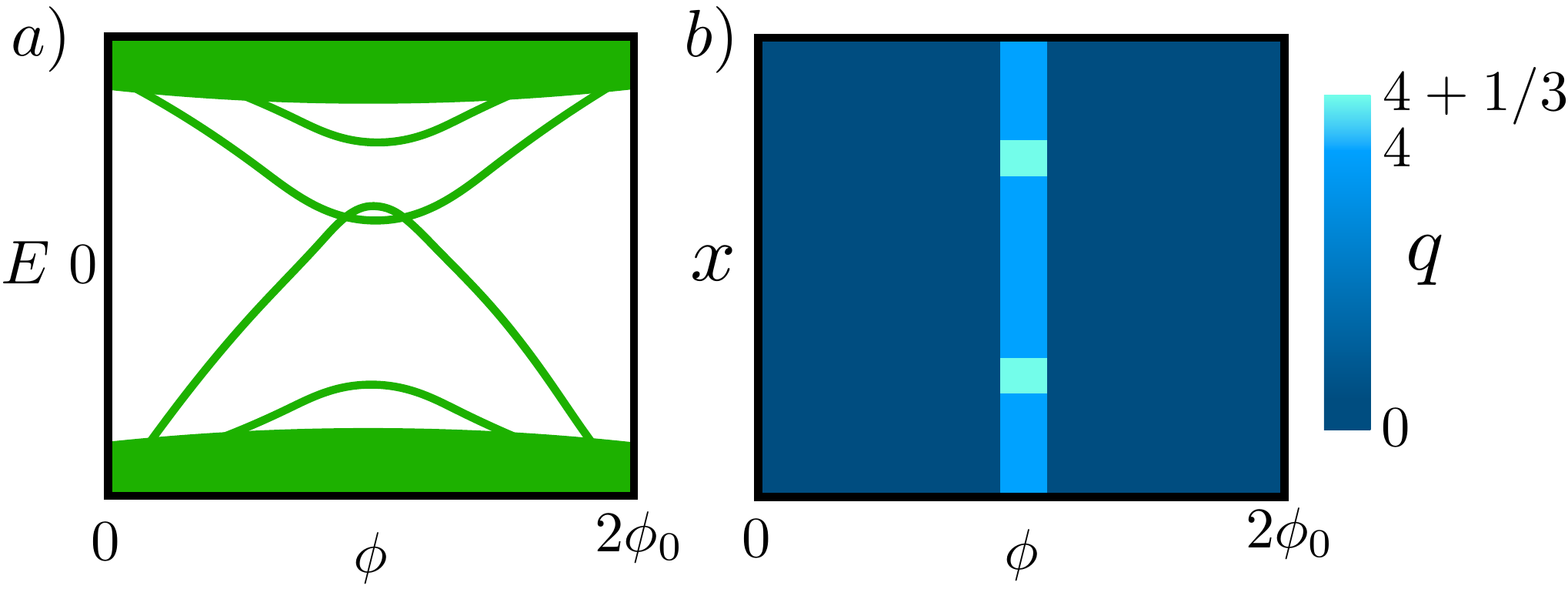}
\caption{ a) Spectrum $E$ vs. flux $\phi$ for the MCI subjected to TRI insertion of magnetic flux at two sites in the lattice. b) topological charge of the occupied states for a LL\textsubscript{F} vs. flux $\phi$ and real-space coordinate in the $\hat{x}$-direction, $x$, for fixed position in the $\hat{y}$-direction corresponding to the two sites of flux insertion through the lattice.}
\label{figmciflux}
\end{figure}

In addition to the bulk-boundary correspondence of the MCI under weak breaking of the bulk Bloch Hamiltonian tensor product structure, we summarise phenomenology of the MCI response to TRI flux insertion reported in Banerjee~\emph{et al.}~\cite{banerjee2024}. In this work, magnetic flux $\phi$ and $-\phi$ are inserted in a time-reversal symmetric manner by Peierls substitution at two locations in lattice tight-binding model realising the MCI state, which are separated in real-space along the $\hat{x}$-axis, respectively, for the topological phase of the MCI considered in the previous section. A $4 \pi$-periodic feature is observed in the plot of MCI spectrum vs. magnetic flux $\phi$, consisting of two pairs of degenerate eigenstates entering the bulk gap and crossing at two values of $\phi$ in the vicinity of one flux quantum, $\phi_0$, as shown in Fig.~\ref{figmciflux} a). In Fig~\ref{figmciflux} b), we show another key result of Banerjee~\emph{et al.}~\cite{banerjee2024} schematically, of the topological charge for a LL\textsubscript{F} vs. flux $\phi$ and real-space position $x$ in the lattice. The topological charge is computed from the structure factor---a generalisation of a structure constant incorporating projection to the occupied subspace of Hilbert space---of a Lie algebra, for the occupied states at each site in the lattice. This topological invariant is formally introduced and discussed in detail in Section III utilising results of the EFT. Notably, the charge computed for the Lie algebra of the occupied subspace deviates from the algebra of the SU(2) generators of the MCI Hamiltonian by $1/3$ over the interval in $\phi$ between crossing of the in-gap bands as shown in Fig~\ref{figmciflux} a), at the locations of flux insertion in the lattice. We identify this $4 \pi$ response of the MCI to TRI flux insertion with that of a fractional quantum Hall (FQH) state at filling factor $\nu=1/2$ in developing the EFT in Section III, while the non-trivial structure factor with additional $1/3$ contribution is evidence that a more general EFT incorporating LL\textsubscript{F}s is relevant to this system.

That the LL\textsubscript{F} is relevant to understanding of the MCI response is, firstly, consistent with MCI phenomenology for weak breaking of the tensor product structure of the bulk Bloch Hamiltonian: we note that local flux insertion is also expected to weakly break the bulk tensor product structure. The possibility of a LL\textsubscript{F} charge encoded in pspin angular momentum is also useful in understanding of MTPs in terms of an effective CS theory for the fractional quantum Hall states, however, as discussed in Section III, as a special case of the more general CS theories of the QSkHE.

\begin{figure}[t]
\includegraphics[width=0.4\textwidth]{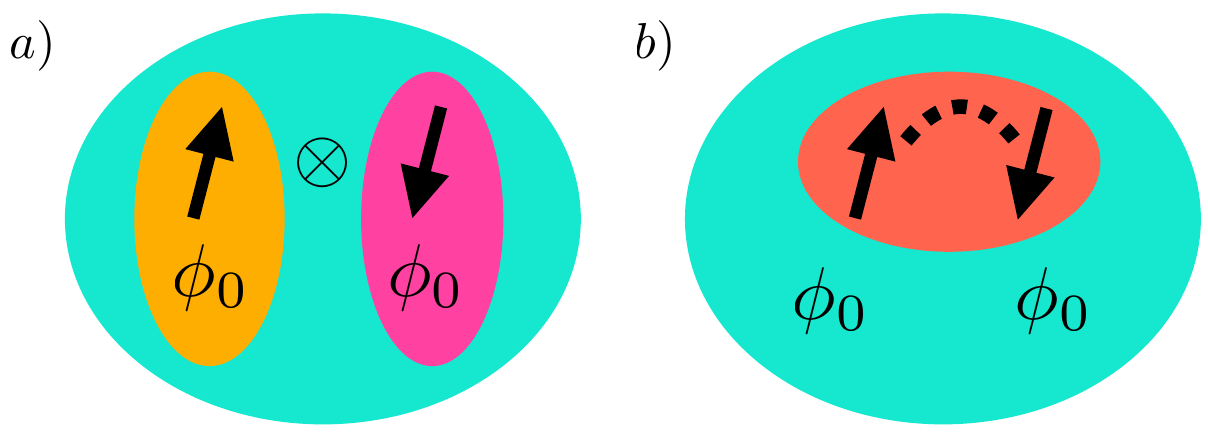}
\caption{a) Composite particle with tensor product structure, schematically depicting effective incompressible parent states, each with one flux quantum per particle. b) Composite particle after breaking tensor product structure. Entanglement between parent particle and hole yield an effective single composite boson, which now sees the two flux quanta, one flux quantum contributed from each parent.}
\label{fig_mci_heuristic}
\end{figure}

Heuristically, realization of a $\nu=1/2$ plateau can also be understood by considering the consequences of breaking the tensor product structure in the system. Interpreting the MCI Hamiltonian as purely quartic in second-quantised creation and annihilation operators, we identify the QP species of the parent Hamiltonians as particles $p$ and holes $h$, respectively. Effective QPs of the MCI Hamiltonian can then be interpreted as $p-h$ pairs. Each $p-h$ pair, which can be thought of as consisting of two particles \textit{isolated} from one another by the tensor product structure, yet forming a composite particle, merge into a single composite boson. The system can then be interpreted as have $N$ such $p-h$ composite bosons and $2N$ flux quanta. This argument is illustrated in Fig.~\ref{fig_mci_heuristic}. We therefore heuristically argue that the system can be understood in terms of more general composite bosons, each consisting of a $p-h$ boson and two flux quanta, corresponding to $\nu=1/2$. We defer discussion of the CS theory to Section III.

\subsubsection{Chirality of real-space spin texture of $\mathcal{C}'$ models through type-II topological phase transition}

Here, we summarize signatures consistent with the interpretation that intrinsically 4+1 D topology is realised in model Hamiltonians with a generalised particle-hole symmetry realising topological skyrmion phases of matter, which are discussed in detail in Winter~\emph{et al.}~\cite{winterOEPT}. The generalised particle-hole symmetry is invariance of the system under the product of a particle-hole conjugation and spatial inversion, or $\mathcal{C}' = \mathcal{C} \mathcal{I}$, where $\mathcal{C}^2=-1$, and therefore relevant to centrosymmetric superconductors~\cite{cook2023}. The basis vector is specified as $\psi^{}_{\boldsymbol{k}} = (c^{}_{\boldsymbol{k},+} c^{}_{\boldsymbol{k},-}, c^{\dagger}_{\boldsymbol{k},-}, c^{\dagger}_{\boldsymbol{k},+})^{\top}$. The Bloch Hamiltonian $H(\boldsymbol{k})$ is a
generalized Bogoliubov de Gennes (BdG) Hamiltonian,
consisting of a generic two band normal state Hamiltonian $H_N(\boldsymbol{k})$
and pairing term $\Delta(\boldsymbol{k})$ as
\begin{equation}
	H_{BdG}(\boldsymbol{k}) = \begin{bmatrix} H_N(\boldsymbol{k}) &
	\Delta(\boldsymbol{k}) \\ \Delta^{\dagger}(\boldsymbol{k}) &
-H^{T}_N(\boldsymbol{k}) \end{bmatrix}
	\label{BdGHam}
.\end{equation}
\\
We consider $H_N(\boldsymbol{k}) = h_{0}(\boldsymbol{k}) \mathbf{I} +
\boldsymbol{h}(\boldsymbol{k}) \cdot \boldsymbol{\sigma} $ and $\Delta(\boldsymbol{k}) = i \Delta_{0} (d_{0}(\boldsymbol{k}) +
\boldsymbol{d}(\boldsymbol{k}) \cdot \boldsymbol{\sigma}) \sigma_{y} $. In
these expressions, $\Delta_{0}$ represents a constant;
$h_{0}(\boldsymbol{k})$ and $d_{0}(\boldsymbol{k})$ are real scalar functions;
$\boldsymbol{h}(\boldsymbol{k})$ and $\boldsymbol{d}(\boldsymbol{k})$ are real
vector functions, $\boldsymbol{\sigma} =
\left( \sigma_x, \sigma_y, \sigma_z \right)$, where $\sigma_{\mu}$ is the
$\mu$\textsuperscript{th} Pauli matrix, and $\mathbf{I}$ is the $2 \times 2$ identity matrix.

We focus on the case in Winter~\emph{et al.}, in which non-trivial skyrmion number $Q$ is realized for trivial total Chern number $C$ at half-filling, corresponding to including constant term
 $H'=\Delta_0 \tau_x \otimes \mathbf{I}_2$ to the Hamiltonian Eq.~\ref{BdGHam}, and taking $\boldsymbol{d}(\boldsymbol{k}) = \left(-\sin(k_y) ,\sin(k_x),0 \right)$, with $\boldsymbol{d}_0(\boldsymbol{k})$ and $\boldsymbol{h}_0(\boldsymbol{k})$ trivial. Such a term respects $\mathcal{C}'$ symmetry but allows for further decoupling of $\mc{Q}$ and $C$ yielding regions of phase space with trivial $C$ and non-trivial $\mc{Q}$.

\subsubsection{Bulk-boundary correspondence of the spin subsystem of the $\mathcal{C}'$ tight-binding model}

\begin{figure}[t]
\includegraphics[width=0.4\textwidth]{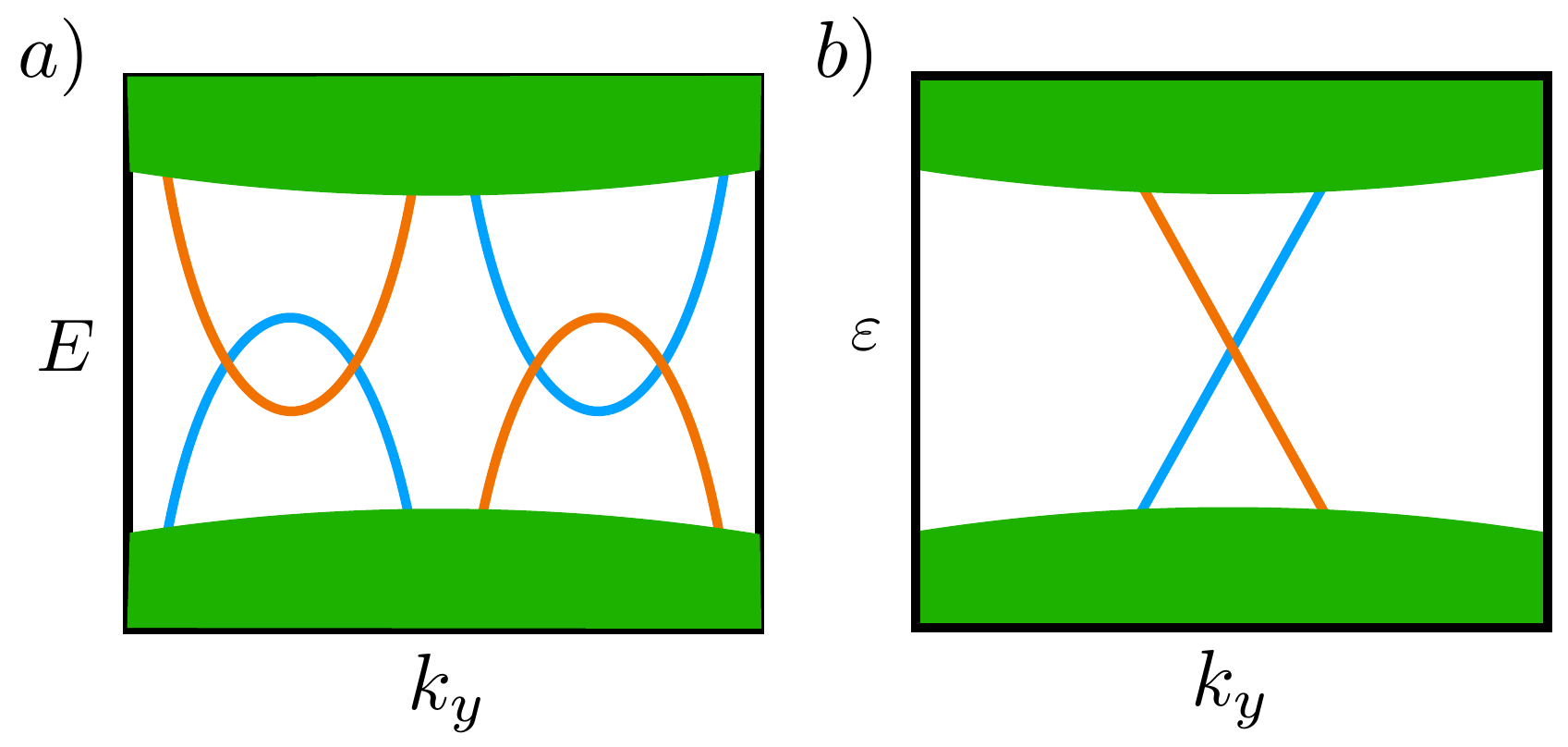}
\caption{Schematic of a) energy slab spectrum for OBC in the $\hat{x}$-direction, PBC in the $\hat{y}$-direction for the $\mathcal{C}'$-invariant model for spin skyrmion number $Q$ non-zero and total Chern number $C$ zero, b) observable-enriched entanglement spectrum for $\mathcal{C}'$-invariant model on a torus with additional real-space virtual cuts for the same parameter set as in (a), with number of chiral modes on each entanglement cut in correspondence with $Q$. Green corresponds to bulk states, purple corresponds to degenerate bands localised on opposite edges, and blue/orange indicate localisation of states on opposite edges.}
\label{fig_OEE_Cp}
\end{figure}

We first discuss results on bulk-boundary correspondence of the $\mathcal{C}'$ model $H_{BdG}(\boldsymbol{k})$ Eq.~\ref{BdGHam}. Similarly to the case of the BHZ model, Winter~\emph{et al.}  examine both the slab energy spectrum of the full four-band model for OBCs in one direction, as well as the OEES, for a system with skyrmion number $\mathcal{Q}=1$ and total Chern number $C=0$. The slab energy spectrum is depicted schematically in Fig.~\ref{fig_OEE_Cp} a). For the same parameter set, the OEES, after tracing out the particle-hole DOF and half of the system in a torus geometry over real-space, is shown in Fig.~\ref{fig_OEE_Cp} b). The slab energy spectrum possesses edge states in the bulk gap similar to those of the MCI when the tensor product structure of the MCI bulk Bloch Hamiltonian is broken. In the case of $H_{BdG}(\boldsymbol{k})$, however, these edge states appear only for OBCs in the $\hat{x}$-direction. A pair of edge bands, consisting of states localised on opposite edges of the system, cross at two points for $k_y<0$, and a counterpart pair of edge bands also exhibits these features, but with localisation reversed relative to the first pair.  After the OEPT, the skyrmion number $\mathcal{Q}$ plays the role of the Chern number of the spin subsystem, and $Q$ chiral modes appear localized near each virtual cut in correspondence with the bulk spin invariant as shown in Fig.~\ref{fig_OEE_Cp} b). 

The topology of the full system is therefore understood, at a certain level, in terms of an intrinsically 2+1 D CS theory for the spin subsystem. The OEPT corresponds, in effect, to tracing out the internal DOFs of the LL QPs encoded in terms of that pspin, and visualizing the bulk-boundary correspondence directly in units of strictly 0 D charged QPs. Before performing the OEPT, however, we interpret these charged QPs of the spin subsystem as LL\textsubscript{F}s encoded in terms of spin angular momentum, encoding two fuzzy spatial dimensions. In combination with the two Cartesian spatial dimensions, the overall dimensionality of the full Hamiltonian is consistent with potentially intrinsically 4+1 D topological states. This is consistent with the bulk-boundary correspondence observed for the slab energy spectrum, which possesses edge bands similar to those of the MCI under symmetry-breaking.

\begin{figure}[t]
\includegraphics[width=0.4\textwidth]{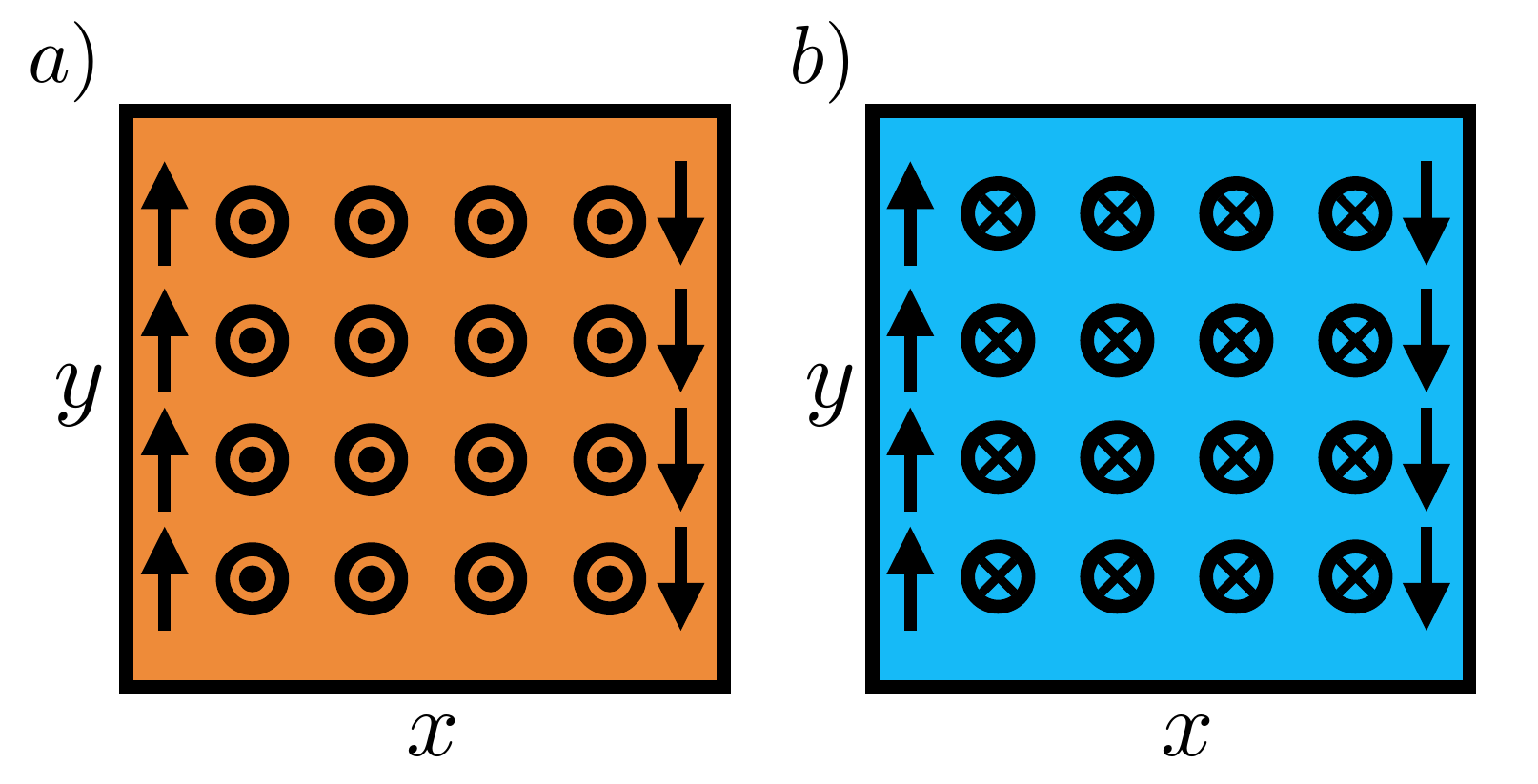}
\caption{Schematics of a) real-space pspin texture for $\mc{Q}=1$ and f) real-space pspin texture for $\mc{Q}=-1$, after passing through a type-II topological phase transition starting with parameter set in a). Orange and blue, along with symbols in system bulk, indicate change in sign of out-of-plane spin component with change in sign of $\mc{Q}$.}
\label{fig_cp_rlspin}
\end{figure}

Similarly to the case of the BHZ model in work by Ay~\emph{et al.}~\cite{ay2024}, Winter~\emph{et al.}~\cite{winterOEPT} also examined real-space pspin textures for occupied states across a type-II topological phase transition, for OBCs in each of the Cartesian spatial directions. A schematic of these results is shown in Fig.~\ref{fig_cp_rlspin}. Spin polarisation is localised and predominantly polarised along the vertical edges, similarly to the case of the BHZ model, while the BHZ model exhibited such polarisation of pspin along all four edges. Notably, change in sign of $\mathcal{Q}$ in both the case of the BHZ model subjected to weak Zeeman field and that of the $\mathcal{C}'$ model yields a change in sign only for the $z$-component of the local pspin expectation value, while in-plane components do not change in sign. This serves as additional evidence that the signatures observed for the BHZ model subjected to weak Zeeman field are those of topological skyrmion phases, linking the bulk-boundary correspondence of $H_{BHZ}(\boldsymbol{k})$ to that of $H_{MCI}(\boldsymbol{k})$ and $H_{BdG}(\boldsymbol{k})$. This is also additional evidence that the spin skyrmion number is playing the role of an effective second Chern number, which could change in sign as a result of change in sign of the LL\textsubscript{F} topological charges. In Fig.~\ref{schem_4Dqhe_2Dqskhe} b), this corresponds to change in sign of charge of the orange LL\textsubscript{F}s, and no change in sign for the green LL.

\subsection{Second Chern  number after fuzzification}

The spin skyrmion number $\mathcal{Q}$ of 2+1 D topological skyrmion phases of matter is the first Chern number of the spin subsystem~\cite{winterOEPT}. Bulk-boundary correspondences and/or response signatures of the full four-band model Hamiltonians are most consistent with signatures of intrinsically 4+1 D topological states characterised by the second Chern number $C_2$, however, suggesting the spin skyrmion number $\mathcal{Q}$ is effectively playing the role of a second Chern number $C_2$ in the full Hamiltonians. This interpretation is consistent with the concept of LL\textsubscript{F}s as intrinsically 2+1 D, yielding topological states in systems with such LL\textsubscript{F}s and, additionally, two Cartesian spatial coordinates that are potentially intrinsically of dimensionality 4+1 D. As stated earlier in section I, however, previous work computing topological invariants for the case of fuzzy dimensions has yielded quantisation of the invariant only in the non-fuzzy limit~\cite{balachandran2006}, and the question of defining quantised invariants in the case of (extra) fuzzy dimensions is an open question. In Section III, which is focused on the EFT of the QSkHE, however, we employ additional machinery to explore computation of a quantised topological invariant for the LL\textsubscript{F}, and discuss its usage in numerics for the MCI topological response~\cite{banerjee2024}.

\subsection{Summary of Section II}

In this section, we have focused on phenomenology of the recently-identified QSkHE and related topologically non-trivial phases of matter indicating that systems with $d$ Cartesian space coordinates, which are then typically expected 
 to realise topologically non-trivial phases of matter that are intrinsically of dimensionality up to $d$+1, can actually encode $\delta$ additional fuzzy dimensions if they possess pspin DOFs. This is the case even when the pspin is associated with $N\times N$ matrix Lie algebra generators and $N$ is small, corresponding to severe fuzzification as defined in this section. Such small $N$ include cases of pspins such as the spin half DOF or a $t_{2g}$ orbital DOF, meaning that, in situations where a pspin has previously been treated as just a label, it actually encodes $\delta>0$ fuzzy dimensions more generally. These $\delta$ additional fuzzy dimensions factor into the intrinsic dimensionality of topological states possible in such systems. For a system defined on a lattice, $\delta$ can also be encoded in a layer pspin DOF corresponding to finite thickness $L$ in some spatial direction, even when $L$ is far from the thermodynamic regime and finite-size effects are prominent, yielding the finite-size topological phases of the QSkHE~\cite{qskhe, cookFST2023}. Systems with $d$ spatial Cartesian coordinates and such pspin DOFS can then realize topological phases of matter of intrinsic dimensionality up to $d+\delta+1$.  
 
 We have reviewed results for the BHZ model for the QSHI observed in HgTe quantum wells~\cite{QSHI-HgTe-Theory}, the MCI~\cite{banerjee2024, cook_multiplicative_2022}, and four-band Bloch Hamiltonian models relevant to centrosymmetric superconductors~\cite{cook2023}, namely bulk-boundary correspondence and topological response signatures. In each case, the Hamiltonian possesses two Cartesian spatial coordinates, but we identify evidence of some topological states more akin to intrinsically 4+1 D states rather than 2+1 D states expected based on the number of Cartesian coordinates of these systems. These results are consistent with LL projection still encoding intrinsically 2+1 D topology after compactification to nearly 0+1 D and $N$ small in some cases, effectively, yielding LL\textsubscript{F}s. In a NL$\sigma$M, LL projection is associated with non-trivial winding of fields forming a skyrmion~\cite{watanabe2014}, with topological charge encoded in the algebra of the momentum commutator. Existence of LL\textsubscript{F}s is consistent with fuzzification of this NL$\sigma$M yielding a quantum counterpart of this skyrmion.  Such quantum skyrmions are defined in terms of non-commuting momentum and position operators with the momentum commutator additionally encoding the topological charge of the skyrmion. 

Importantly, the $4 \pi$-periodic Aharanov-Bohm effect observed for time-reversal symmetric insertion of magnetic flux for the case of the MCI response provides strong evidence of pairing of LL\textsubscript{F}s associated with different pspin DOFs to form composite LL\textsubscript{F}s by entangling these DOFs. A compactified many-body state associated with filling some number of orbitals of a LL\textsubscript{F} after entangling these pspin DOFs is potentially composite in nature, generalising Jain's notion of a composite QP \cite{Jain:1989tx}. That is, the QSkHE is consistent with a qualitative generalization from single-particle topology, even in lattice tight-binding models previously-assumed to describe only effectively non-interacting systems, such as four-band Bloch Hamiltonians that are quadratic in second-quantized creation and annihilation operators. 

\section{Phenomenological Chern-Simons field theories}

Implications of the phenomenology reviewed in Section II of the manuscript are that topological phases of the QSkHE in $d+1$-dimensions (based on the number of Cartesian space coordinates utilized in the action, $d$), for systems with pspin DOFs, are generically up to intrinsically $d+\delta+1$-dimensional, with $\delta \ge0$, due to the pspin DOFs potentially encoding $\delta$ fuzzy dimensions, even for pspin DOF(s) encoded in terms of Lie algebra generators with matrix representation $N\times N$  and $N$ small (e.g., $N=2$ and $N=4$). Of existing theoretical machinery, this $d+\delta+1$-dimensional topology is best captured by an action, which depends on a combination of Cartesian space coordinates, as well as fuzzy coset space coordinates encoding additional spatial dimensions~\cite{aschieri2007}.

In this second half of the manuscript, we unify insights from Section II on phenomenology to construct effective field theories capturing the QSkHE, within this framework of extra fuzzy dimensions. We first review relevant gauge theories in subsection A and then introduce effective Chern-Simons (CS) theories for the QSkHE starting with a continuum counterpart of the MCI in Subsection B and ending with an EFT of the QSkHE within the fuzzy extra dimensions framework~\cite{aschieri2007}. In subsection C, we generalise beyond the fuzzy extra dimensions framework by considering base manifolds with local product structure, and introduce methods for computing quantised topological charge of LL\textsubscript{F}s. We also finally discuss 6+1 D gauge theories related to the 4+1 D SU(2) gauge theory, which are helpful in understanding some of the more complex phenomenology of the lattice tight-binding models, in particular the rich interplay between bulk-boundary correspondence and topological response signatures.

\subsection{Gauge theories associated with the second Hopf map}

A 4+1 D generalisation of the quantum Hall effect (QHE) has been proposed in \cite{zhanghu2001} employing the second Hopf Map $S^7\to S^4$. The theory considers an SU(2) monopole at the centre of the four-sphere $S^4$, where the particles are carrying SU(2) isospin index $I$. 

The second Hopf map is given by the equation,
\begin{equation}
    \frac{X_{\mu}}{R}=\Bar{\Psi}\Gamma_{\mu}\Psi,
    \label{2ndHopfmap}
\end{equation}
where $\Psi$ is a normalised, four-component complex spinor (so $\Bar{\Psi}\Psi=1$), the radius $R$ of four-sphere $S^4$  satisfies $\sum_{\mu}X_{\mu}X^{\mu}=R^2$ and $\Gamma_{\mu}$ are five Dirac matrices satisfying the Clifford algebra. The non-Abelian $SU(2)$ holonomy over this map is given by \cite{Demler:1998pm},
\begin{equation}
    A_{\mu}=\frac{-i}{R(R+X_5)}\eta^{i}_{\mu\nu}I_iX_{\nu},\qquad X_5=0,
\end{equation}
where here $\eta^{i}_{\mu\nu}=\epsilon_{i\mu\nu4}+\delta_{i\mu}\delta_{4\nu}-\delta_{i\nu}\delta_{4\mu}$ is the 't Hooft symbol and $I_i$ are isospin $SU(2)$ generators in representation $I$. It is also the Yang monopole potential~\cite{Yang:1977qv}. The single particle Hamiltonian describing the particle moving in the monopole background as,
\begin{equation}
    H=\frac{\hbar^2}{2MR^2}\Lambda^2_{\mu\nu}
\end{equation}
where, $\Lambda_{\mu\nu}=-i (X_{\mu}D_{\nu}-X_{\nu}D_{\mu})$ are covariant angular momentum operators. The lowest Landau level (LLL) is $\frac{1}{6}(p+1)(p+2)(p+3)$-fold degenerate with $p=2I$. The corresponding wavefunctions are,
\begin{equation}
    \braket{X_a|m_1,m_2,m_3,m_4}=\mathcal{N}_m\Psi_1^{m_1}\Psi_2^{m_2}\Psi_3^{m_3}\Psi_4^{m_4}
\end{equation}
with $m_1+m_2+m_3+m_4=p$ and $\mathcal{N}_m\sqrt{\frac{p!}{m_1!m_2!m_3!m_4!}}$. 

A notable feature of the theory is non-commutative geometry defined by the relation,
\begin{equation}
    [X_{\mu},X_{\nu}]=4il^2\eta^i_{\mu\nu}\frac{I_i}{I}
\end{equation}
with the parameter $\lim\limits_{R,p \to \infty} R^2/p=l^2$ parameterising the non-commutativity.

Boundary excitations of the 4+1 D QHE are notable for including higher helicity collective modes~\cite{Zhang:2001xs, Hu:2001kf, elvang2003quantum}. Considering a compact manifold such as $S^4$, which does not incorporate a boundary, the boundary excitations are realised by introducing a confining potential $V(X_5)$. This confining potential reduces the $SO(5)$ symmetry of the model to $SO(4)\simeq \left(SU(2)\times SU(2) \right)/\mathbb{Z}_2$ symmetry.  The collective excitations of the quantum Hall liquid are then particle-hole excitations labelled by the  Casimir operators of this semisimple group. The collective bosonic modes are labeled in terms of helicity $S$ satisfying $0\leq S\leq I$, where $I$ defines the irreducible representation of $SU(2)$ monopole. \\

At this point, we return to comparison of the 4+1 D CI and the 4+1 D QHE to highlight an intriguing aspect of the 4+1 D QHE theory. Most importantly, there exist no-go theorems, such as the Weinberg-Witten theorem~\cite{Weinberg:1980kq}, which mean composite higher-spin particles are forbidden in some field theories. In the EFT of the 4+1 D QHE, however, there are graviton-like, spin $2$ modes and even higher-spin modes, which is an unphysical ``embarrassment of riches'' of the theory absent in models for the 4+1 D CI. These riches triggered a flurry of interest from the high-energy physics community in the 4+1 D QHE, given the existence of the no-go theorems. We will return to this point after introducing the minimal EFT of the QSkHE as that of the 4+1 D SU(2) gauge theory subjected to generalised fuzzification consistent with phenomenology of Section I.\\

In previous work~\cite{bernevig6Dfieldtheory}, gauge theories associated with the second Hopf map, such as the 4+1 D SU(2) gauge theory of greatest relevance to the present work, are derived starting first from a 6+1 D U(1) gauge theory, which we review in greater detail in a later section. For the time being, we are focused on the 4+1 D SU(2) gauge theory, and simply note that this theory has previously been derived from the 6+1 D $U(1)$ Abelian CS theory over configuration space $CP^3$ given by
\begin{equation}
    S=\nu\int d^7x\; A\wedge dA\wedge dA,
\end{equation}
by simplified dimensional reduction and fuzzification, which assumed fixed expectation values for NL$\sigma$M field components over the two Cartesian coordinates to be removed as part of dimensional reduction~\cite{bernevig6Dfieldtheory}. This 6+1 D U(1) theory is reduced to a 4+1 D $SU(2)$ CS theory with the Lagrange density,
\begin{align}
\mathcal{L} &= \frac{4\pi \nu}{3} \mathrm{Tr} \left( \boldsymbol{A} \wedge d\boldsymbol{A} \wedge d\boldsymbol{A}  - {3 i \over 2} \boldsymbol{A} \wedge \boldsymbol{A} \wedge \boldsymbol{A} \wedge d\boldsymbol{A} \right. \\ \nonumber
&\left. - {3 \over 5} \boldsymbol{A} \wedge \boldsymbol{A} \wedge \boldsymbol{A} \wedge \boldsymbol{A} \wedge \boldsymbol{A}  \right) .
\end{align}\\
One can identify the coefficients appearing in front of the action with second Chern number $C_2$ and identify the topological response with this topological invariant. Although we first discuss simpler CS theories for the QSkHE motivated specifically by phenomenology of the MCI, we later return to this Lagrange density and apply generalized fuzzification to it, to arrive at the EFT of the QSkHE defined over 2+1 D Cartesian coordinates. We will show that, under generalized fuzzification consistent with lattice model phenomenology summarized in Section I, this EFT of the QSkHE is distinguished from the established 2+1 D SU(2) gauge theory by higher-dimensional terms such as those 4+1 D terms of the above Lagrange density. These additional terms yield topology more akin to that of intrinsically 4+1 D topological states, for instance, than 2+1 D, as a result of two spatial dimensions being encoded in pspin DOFs resulting from the generalized fuzzification procedure.

\subsubsection{Spinor of the second Hopf map reduced to multiplicative case}
Before discussing generalized fuzzification of the 4+1 D SU(2) gauge theory, we first briefly consider the spinor $\Psi$ of the 4+1 D SU(2) gauge theory with additional symmetry protection of MTPs, to reduce $\Psi$ to a form relevant to these topological states.

We first define $\Psi$ as the following,
\begin{equation}
    \Psi = \begin{pmatrix}
        \psi_1\\
        \psi_2\\
    \end{pmatrix}\otimes\begin{pmatrix}
        \phi_1\\
        \phi_2\\
    \end{pmatrix}=\begin{pmatrix}
        \psi_1\phi_1\\
        \psi_1\phi_2\\
        \psi_2\phi_1\\
        \psi_2\phi_2\\
    \end{pmatrix}
\end{equation}
such that $\Psi^{\dagger}=(\phi_1^{\dagger}\psi_1^{\dagger}\; \phi_2^{\dagger}\psi_1^{\dagger}\; \phi_1^{\dagger}\psi_2^{\dagger}\; \phi_2^{\dagger}\psi_2^{\dagger})$.  With the assumption, $[\psi_{\alpha},\;\phi_{\beta}]=0$ one finds,
\begin{align}
    \Psi^{\dagger}d\Psi&=(\phi_1^{\dagger}\psi_1^{\dagger}\; \phi_2^{\dagger}\psi_1^{\dagger}\; \phi_1^{\dagger}\psi_2^{\dagger}\; \phi_2^{\dagger}\psi_2^{\dagger})\begin{pmatrix}
        d\psi_1\phi_1+\psi_1 d\phi_1\\
        d\psi_1\phi_2 +\psi_1 d\phi_2\\
        d\psi_2\phi_1+\psi_2 d\phi_1\\
        d\psi_2\phi_2+\psi_2d\phi_2\\
    \end{pmatrix} \\ \nonumber&=\psi^{\dagger}d\psi+\phi^{\dagger}d\phi
\end{align}
Using the second Hopf Map, $n_i=\Psi^{\dagger}\Gamma_i\Psi$ one gets,
\begin{align}
    n_i&=-(\psi^{\dagger}\sigma_2\psi)(\phi^{\dagger}\tau_i\phi), \qquad i=1,2,3\\
    n_4&=(\psi^{\dagger}\sigma_1\psi)(\phi^{\dagger}\mathbf{I}\phi)\\
    n_5&=(\psi^{\dagger}\sigma_3\psi)(\phi^{\dagger}\mathbf{I}\phi)
\end{align}
where $\mathbf{I}$ is the identity matrix. Similar to past work~\cite{Nakayama:2004vm}, using the first Hopf Maps $x_i=\psi^{\dagger}\sigma_i\psi$ and $y_i=\phi^{\dagger}\tau_i\phi$, we can write,
\begin{align}
    n_i&=-x_2 y_i, \qquad i=1,2,3\\
    n_4&=x_1\\
    n_5&=x_3.
\end{align}
Again promoting $(x_i,y_j)$ to respective Pauli matrices, one can see the $n_i$'s satisfy the Clifford algebra $\{n_i,n_j\}=\delta_{ij}$. The Hamiltonian \cite{Demler:1998pm} in this tensor product form becomes,
\begin{equation}
    \mathcal{H}= n_i\Gamma^i=-x_2y_i(\sigma^2\otimes \tau^i)+(x_1\sigma^1+x_3\sigma^3)\otimes \mathbf{I}
\end{equation}
where we used the tensor product form of Gamma matrices \cite{Zhang:2001xs}. The tensor product spinor defines a map as,
\begin{equation}
   S^3\times S^3 \xrightarrow{U(1)\times U(1)} S^2\times S^2.
\end{equation}
The $S^2\times S^2$ here is the double cover of $S^4$ have a singularity along $x_2$. In terms of the second Hopf Map, $\Psi$ is an eigenstate of the above Hamiltonian, and one can show,
\begin{align}
    H\Psi &=n_i\Gamma^i\Psi=\Psi^{\dagger}\Gamma_i\Psi\Gamma^i\Psi=c_i\Phi^i\\
    \Psi^{\dagger}H\Psi &=(\Psi^{\dagger}\Gamma_i\Psi)(\Psi^{\dagger}\Gamma^i\Psi)=c_i\Psi^{\dagger}\Phi^i\\
    &=\sum_i n_i^2=1= c_i\Psi^{\dagger}\Phi^i
\end{align}
where the trivial solution is $\Phi^i=\Psi$ and $c_i=1$.  \\

While the above discussion clearly shows the product structure of the spinor associated with the symmetry-protected tensor product structure of Hamiltonians realising MTPs of matter, we note that the Grassmannian for the MTPs relevant to four-band models is SO(4)/(SO(2)$\times$SO(2))~\cite{cook_multiplicative_2022}. This is a five-dimensional quadric surface of CP$_3$~\cite{arai2007, KLEIN200879}. The Lagrange density for the MTPs in greater generality is then a 5+1 D U(1)$\times$U(1) theory. In addition to the case of gauge theories for particles on $S^4$ seeing an SU(2) monopole ($S^3$)  associated with the second Hopf map, a six-sphere $S^6$ and U(1) monopole ($S^1$) theory has also been considered previously~\cite{bernevig6Dfieldtheory} and will be discussed in greater detail in the final subsection of the manuscript. The MTPs correspond, more generally, to a previously-unconsidered intermediate case of five-sphere $S^5$ and U(1)$\times$ U(1) monopole ($S^2$), which also corresponds to the second Hopf map $S^7 \rightarrow S^4$.

\subsection{Minimal Chern-Simons field theories starting from multiplicative topological phases}

Consider two $2+1$ D systems of particles related by TRS, labeled $1$ and $2$. Let system $1$ realize a CI state with $C=1$, and system $2$ realize a CI state with $C=-1$, as in Banerjee~\emph{et al.}~\cite{banerjee2024}. Let the particles of system $2$ be holes of system $1$. Then system $1$ corresponds to an incompressible state of $N$ particles and $N$ flux quanta $\phi_0$, and system $2$ corresponds to $N$ holes and $N$ flux quanta $\phi_0$. Specifically, let us consider the particles to be spinless fermions.

We first consider the limit of very strong correlations forming $p-h$ pairs. In the case of spinless fermion particles, these pairs constitute spin singlets. This corresponds to a child Bloch Hamiltonian for the MCI of $\mc{H}_c(\boldsymbol{k}) = \mc{H}_1(\boldsymbol{k}) \otimes \mc{H}_2(\boldsymbol{k})$, which can describe a symmetry-protected multiplicative topological phase~\cite{cook_multiplicative_2022}. Eigenstates of the Bloch Hamiltonian are at least two-fold degenerate for each $\boldsymbol{k}$, yet the symmetry-protected tensor product structure yields a $U(1) \times U(1)$ holonomy. 

Given this holonomy, we initially consider commutative theories to simplify the discussion, and will later generalize to non-commutative theories when discussing generalised fuzzification from 4+1 D to 2+1 D.

A consequence of the possibility of severely-fuzzified Landau levels, LL\textsubscript{F}s, is the notion of charge encoded in a pspin DOF as the topological invariant of a compactified many-body state. Such a pspin charge---and corresponding gauge field---can be defined for the composite particle.  For the MCI, we therefore consider a minimal effective action with three terms:

\begin{align}
S_{p,\mathrm{eff}} &= {C_p \over 4 \pi} \int  d^2x dt \varepsilon^{\mu \nu \rho} A_{p,\mu} \partial_{\nu} A_{p, \rho}\\
S_{h,\mathrm{eff}} &=  {C_h \over 4 \pi} \int  d^2x  dt 
 \varepsilon^{\mu \nu \rho} A_{h,\mu} \partial_{\nu} A_{h, \rho} \\
S_{c,\mathrm{eff}} & = {C_p\over 2 \pi} \int  d^2x  dt 
 \varepsilon^{\mu \nu \rho} A_{p/h,\mu} \partial_{\nu} a_{\rho}\\ \nonumber
 &+{Q\over 4 \pi} \int  d^2x  dt 
 \varepsilon^{\mu \nu \rho} a_{\mu} \partial_{\nu} a_{\rho},
\end{align}

where we write $\partial_{\nu} a_{\rho}$ to make explicit that the current density of the composite particles is an effective emergent current density associated with the composite particles carrying charge encoded in terms of a pspin, which consolidates spin contributions from the parent particle species. $A_{p/h,\mu} = A_{p,\mu} + A_{h,\mu}  $ and $a_{\mu} = A_{p,\mu} + T\left[A_{h,\mu}\right]$ are U(1) $\times$ U(1) gauge fields. $A_{p/h,\mu}$ acts on the composite particles by the $A_{p,\mu}$ component acting on the parent $1$ particle species and the $A_{h,\mu}$ component acting on the parent $2$ particle species. $T\left[A_{h,\mu}\right]$ denotes performing a particle-hole transformation on $A_{h,\mu}$, as required to consolidate gauge field contributions from systems $1$ and $2$ into a single gauge field seen by the composite particles specifically due to their pspin charge. $a_{\mu}$ is therefore the emergent gauge field encoded in pspin angular momentum and associated specifically with the charge of the composite particle encoded in pspin angular momentum in general.  

Since the composite particles consist of $p-h$ pairs, $a_{\mu} = 2 A_{p,\mu}$, yielding $Q=2$ for $C_p=1$. In this case, the effective gauge theory possesses the CS terms for the $\nu = 1/2$ plateau. This is consistent with the $4 \pi$ Aharonov-Bohm effect observed for such a MCI with two CI parents related by TRS, one parent possessing Chern number $C=1$ and the other with Chern number $C=-1$, as discussed in Section I. $Q$ can therefore be interpreted as an invariant associated with a term of a 2+1 D CS theory for the fractional quantum Hall effect (FQHE) with filling factor $\nu = 1/2$.

However, we can also interpret the pspin gauge field $a_{\mu}$ as being that seen by LL\textsubscript{F}s given the charge is encoded in the pspin DOFs, to generalise this theory. Indeed, if the tensor product structure is broken while retaining particle-hole symmetry, say by local flux insertion, the $p-h$ pair is expected to generalise to a LL\textsubscript{F}. 

If we generalize by interpreting the pspin charge as associated with LL\textsubscript{F}s, we may first generalise the theory to 4+1 dimensions,  assuming the LL\textsubscript{F} encodes two fuzzy extra dimensions. We focus here on the
most essential terms for our discussion of topologically non-trivial states in the lattice models and specifically intrinsically
4+1 D terms.

\begin{align}
S_{p,\mathrm{eff}} &= {C_p \over 4 \pi} \int  d^2x dt \varepsilon^{\mu \nu \rho} A_{p,\mu} \partial_{\nu} A_{p, \rho}\\
S_{h,\mathrm{eff}} &=  {C_h \over 4 \pi} \int  d^2x'  dt 
 \varepsilon^{\mu \nu \rho} A_{h,\mu} \partial_{\nu} A_{h, \rho} \\
S_{c,\mathrm{eff}} & = {C_p \over 12 \pi^2} \int d^4x dt \varepsilon^{\mu \nu \rho \sigma \tau} A_{p/h,\mu} \partial_{\nu} \left( a_{\rho}  \partial_{\sigma} a_{\tau} \right)\\
&+{Q \over 24 \pi^2} \int d^4x dt \varepsilon^{\mu \nu \rho \sigma \tau} a_{\mu} \partial_{\nu} \left( a_{\rho}  \partial_{\sigma} a_{\tau} \right).
\end{align}

That is, the current density generalizes from $\partial_{\nu} a_{\rho}$ to  $\partial_{\nu}\left( a_{\rho}  \partial_{\sigma} a_{\tau} \right) $. We note that $\partial_{\nu} \left( a_{\rho}  \partial_{\sigma} a_{\tau} \right) = \partial_{\nu} a_{\rho}  \partial_{\sigma} a_{\tau}$. 

While this Abelian 4+1 D theory streamlined discussion, we now consider the intrinsically 4+1 D terms of the SU(2) gauge theory. We apply a dimensional reduction procedure to the non-Abelian 4+1 D SU(2) gauge theory, of severe fuzzification, yielding 2+1 D SU(2) gauge theory given the remaining number of Cartesian spatial coordinates. However, this 2+1 D SU(2) gauge theory, in agreement with phenomenology, contains  additional terms absent in the established 2+1 D SU(2) gauge theory~\cite{affleck1988, Demler:1998pm}, which are vestiges of the underlying 4+1 D theory.

Fuzzification for the purpose of dimensional reduction is considered in past work~\cite{Kapetanakis:1992hf, Forgacs:1979zs, Aschieri:2003vy, bernevig6Dfieldtheory}. In the present work, however, we are guided by phenomenology of the lattice models~\cite{qskhe,cook_multiplicative_2022, cookFST2023}, and therefore employ fuzzification to remove two Cartesian spatial coordinates specifically by performing severe fuzzification \textit{while retaining  dependence of the field theory on these severely-fuzzified coordinates}. This yields an EFT that appears to be 2+1 D based on its Cartesian coordinates, but is more akin to the 4+1 D theory based on the intrinsically 4+1 D forms of previously-unconsidered, additional terms in the Lagrange density. 

We will arrive at a CS theory defined over two remaining Cartesian spatial coordinates, $x$ and $y$, with the two other Cartesian coordinates, $z$ and $w$, replaced by coset space coordinates of a fuzzy two-sphere $S^2_F$. Such fuzzy ``extra'' dimensions have been considered previously~\cite{aschieri2007}. Using notation of this and related works \cite{Aschieri:2003vy, Aschieri:2004vh,Chatzistavrakidis:2010tq}, we 
reduce from the 4+1 D SU(2) gauge theory to a gauge theory defined over 2+1 D Cartesian coordinates, or over $M^{2+1}$, as well as fuzzy coset space coordinates of $(S/R)_F$. The theory is then overall defined over $M^{2+1} \times (S/R)_F$. For the present work, we will focus on the case of $(S/R)_F = \left(SU(2)/  U(1) \right)_F $ to first illustrate the essential points. We note, however, that the coset space most relevant to understanding of the topological skyrmion phases~\cite{cook2023} and MTPs~\cite{cook_multiplicative_2022} in the four-band Bloch Hamiltonian tight-binding models is actually $(S/R)_F = \left( SO(5)/  SU(2)\right)_F  $. This is the coset space associated with the EFT of the 4+1 D SU(2) QHE and the corresponding high-symmetry 4+1 D CI.

\subsubsection{Calculus over fuzzy spaces}
One defines the fuzzy sphere through the embedding of co-ordinates into $N \times N$ matrix representations of $SU(2)$ generators through \cite{Chatzistavrakidis:2010tq}
\begin{equation}
    \left[X_a, X_b\right]=C^c_{ab}X_c
\end{equation}
The differential calculus over the fuzzy sphere can be defined by assuming the action of derivatives by Lie brackets,
\begin{equation}
    e_a(f)=[X_a,f], \qquad df=[X_a,f]\theta^a
\end{equation}
where $e_a$ is derivation in the $X_a$ direction over matrix valued function $f$. $d$ is the exterior derivative. The cotangent space can be generated by the basis $\theta^a$'s, so that the one form over the manifold can be written as,
\begin{equation}
    A=A_{\mu}dx^{\mu}+A_a\theta^a.
\end{equation}

Lastly, $k\Tr$ denotes integration over the fuzzy coset space $(S/R)_F$ described by $N \times N$ matrices, where $k$ is a parameter related to the size of the fuzzy coset space~\cite{aschieri2007}. In this work, we will ultimately consider very small $N$ of $2$ and $4$ but maintain generality for now. 

\subsubsection{Gauge theory over fuzzy spaces}
For a field over the fuzzy sphere, $\phi(X_a)$, the infinitesimal gauge transformation is given as,
\begin{equation}
    \delta_{\lambda}\phi=\lambda(X)\phi
\end{equation}
where the gauge transformation parameter is valued in the co-ordinates Lie algebra, while it can be group $G$ valued in itself. We note that the commutator $[X_a,\lambda]$ might not be zero. One assumes that the co-ordinates are invariant under the gauge transformation such that $\delta_{\lambda}X=0$. One then introduces covariant derivatives/co-ordinates as
\begin{equation}
    \phi_a=X_a+A_a.
\end{equation}
The covariance condition is $\delta\phi_a=[\lambda, \phi_a]$. Due to the nonzero overlap between co-ordinates, the transformation $\delta A_a=[\lambda, X_a]+[\lambda, A_a]$ follows. From the connection, one can define the field strength,
\begin{align}
    F_{ab}&=\left[X_a,A_b\right]-\left[X_b,A_a\right]+\left[A_a,A_b\right]-C^c_{ab}A_c.
\end{align}
One can check that for a gauge covariant condition $\delta F_{ab}=\left[\lambda,F_{ab}\right]$, one should include non zero $[A_a, A_b]$ and $C_{ab}^cA_c$ terms while defining the field tensor. We note that it is possible, even if one initially considers an Abelian gauge group for a $d +\delta+1$ D action, with $d+\delta$ Cartesian space coordinates, to obtain a non-Abelian
gauge symmetry via fuzzification to $d$ Cartesian space coordinates and $\delta$ fuzzy coset space coordinates \cite{Aschieri:2003vy}. 

Following Aschieri~\emph{et al.}~\cite{aschieri2007}, we decompose the field strength $F_{MN}$ in terms of a component defined over the remaining Cartesian coordinates as $F_{\mu \nu}$, a component defined over the fuzzy sphere as $F_{ab}$, and a mixed term $F_{\mu a}$, which take the following forms:
\begin{align}
    F_{\mu \nu} & = \partial_{\mu} A_{\nu} - \partial_{\nu} A_{\mu} + \left[ A_{\mu}, A_{\nu} \right], \\ \nonumber
     F_{ab} &= \left[ X_a, A_b \right]   - \left[ X_b, A_a \right] + \left[A_a, A_b \right] - C^c_{ab} A_c \\ \nonumber
     F_{\mu a} &= \partial_{\mu} A_a  - \left[ X_a, A_{\mu} \right] + \left[A_{\mu}, A_a \right]
\end{align}
In terms of covariant co-ordinates $\Phi=X_a+A_a$, we get a simplified form,
\begin{align}
    F_{\mu a}&=D_{\mu}\phi_a\\
    F_{ab}&=\left[\phi_a,\phi_b\right]-C^c_{ab}A_c.
\end{align}
 
For the present purposes, we apply this machinery to a simplification of the 4+1 D SU(2) gauge theory for the QHE. We consider the flat space limit of this theory~\cite{bernevig6Dfieldtheory}, in which the theory decomposes into a theory of two 2+1 D QHEs, such as over the $xy$ plane and the $zw$ plane. We may then write the gauge theory of the 4+1 D QHE as an Abelian $U(1)$ CS action over $M^{1,2}\times S^2$,
\begin{equation}
    S_{CS}=C_2\int d^5x\; AdAdA, 
\end{equation}
where here the second Chern number $C_2$ is a product of two first Chern numbers, one for each 2+1 D QHE, or $C_{xy}$ and $C_{zw}$, respectively. That is, $C_2 = C_{xy} C_{zw}$.

This theory possesses a product structure amenable to fuzzifying two spatial dimensions within the framework of extra fuzzy dimensions. We fuzzify $S^2$, yielding the action for the theory in terms of a mixture of two Cartesian spatial coordinates and two extra fuzzy dimensions as,
\begin{align}
    S_{CS}=&C_2\int d^3x k\;Tr\left[Tr(G)\;CS_3\wedge \Hat{F}\right]\\
   CS_3=&\left[AdA+\frac{2}{3}AAA\right]\\
   \Hat{F}= & \left[ X_a, A_b \right]   - \left[ X_b, A_a \right] + \left[A_a, A_b \right] - C^c_{ab} A_c,
\end{align}
yielding an action similar in form to a strictly 2+1 D CS action save for a particular matrix structure. A similar form is considered in the context of Kähler Chern-Simons theories \cite{Nair:1990aa}.

We can simplify the discussion to some degree similarly to Qi~\emph{et al.}~\cite{qi2008TRIFT}, by focusing on the response signature specifically of a 4+1 D band topological insulator with single particle Hamiltonian $h(k)$. The non-linear response coefficient is the second Chern number, $C_2$. To gain insight into the phenomenology of the QSkHE, we may write the current density associated with topological response of the 4+1 D QHE and CI as~\cite{qi2008TRIFT},
\begin{align}
    J^L =3 {C_2} \epsilon^{LMNOP}F_{MN} F_{OP}.
\end{align}
Considering we made the decomposition $\epsilon^{LMNOP}=\epsilon^{\mu\nu\rho ab}$ as part of fuzzifying within the framework of extra fuzzy dimensions, one can more specifically write
\begin{align}
    J^{\mu}=3C_2 \epsilon^{\mu\nu\rho ab}F_{\nu\rho}F_{ab},
\end{align} 
for, in effect, both the 4+1 D QHE and 2+1 D QSkHE within the extra fuzzy dimensions framework. 

We see the current is proportional to the flux $F_{ab}$ passing through the fuzzy sphere, associated with LL\textsubscript{F}s. That is, we can understand the phenomenology of the QSkHE from the perspective of extra fuzzy dimensions, as that of the 4+1 D QHE in the flat space limit, substituting LLs for LL\textsubscript{F}s. That is, in the case of the 4+1 D QHE, response to external fields yields topological transport of LLs, while response to external fields in the 2+1 D QSkHE yields topological transport of LL\textsubscript{F}s. Similarly, in the case of the 4+1 D QHE, bulk-boundary correspondence may be understood in terms of LLs propagating on the boundary, while the bulk-boundary correspondence in the 2+1 D QSkHE may be understood in terms of LL\textsubscript{F}s propagating on the boundary. Such effectively higher-dimensional topological response signatures due to a mix of Cartesian and fuzzy coset space coordinates are consistent with phenomenology of Section I. 

We note that the intuitive picture of the QSkHE in terms of severely-fuzzified branes (e.g., LL\textsubscript{F}s) coupling to appropriate fields is broadly-applicable given past work on EFTs for higher-dimensional quantum Hall effects~\cite{Wu:1988py,heckman20186d, Palumbo:2021dkx, Myers:1999ps, Kimura:2004hf, Hasebe:2014nia, Hasebe:2014nia, Hasebe:2003gx}: such EFTs are broadly based on coupling of membranes (e.g., LLs) and higher-dimensional branes to appropriate higher-form fields. Considerable insight into more complex cases of the QSkHE may then come from these EFTs, possibly very directly through conversion of some dimensions encoded in Cartesian space coordinates to extra fuzzy dimensions encoded in fuzzy coset space coordinates, and corresponding fuzzification of these branes and associated fields. However, one important distinction between these higher-dimensional EFTs and those of the QSkHE may be that fuzzification potentially introduces effective spin-orbit coupling---and entanglement---between fuzzy coset space coordinates. The consequences of such entanglement, when isospin DOFs are interpreted as encoding some number of spatial dimensions, is perhaps one of the most important topics of future study. 

The need to consider the flat space limit of the 4+1 D QHE to employ the machinery of extra fuzzy dimensions to formulate the field theory of the 2+1 D QSkHE raises the question of what generalisations are required to capture the theory of the QSkHE without this simplification. We discuss this over the next two sections, first by introducing methods to compute quantised topological invariants over fuzzy spaces based on the phenomenology of Section I.

\subsection{Generalising the effective theory of the quantum skyrmion Hall effect beyond extra fuzzy dimensions}

Having discussed the limitations of the extra fuzzy dimension machinery in its restriction to manifolds with global product structure, we generalise based on the phenomenology of Section I, both by presenting a method for computing quantised topological invariants on fuzzy spaces, and employing this in combination with the dimensional hierarchy of higher-dimensional (fuzzy) spheres in terms of (fuzzy) two-spheres~\cite{Hasebe:2010vp} to construct a more general action of the 2+1 D QSkHE. Within this more general framework, we also discuss interpretation of the intertwining bulk-boundary correspondence and topological response signatures presented in Section I, from a field theory perspective.

\subsubsection{Reformulation of the non-linear sigma model central extension}

For the purposes of discussing topological invariants on fuzzy spaces in the next section, we first reformulate the derivation of the central extension for the quantum ferromagnet and Landau model in Section I, in terms of more general expressions. We first expand the commutator $\left[P^i,T^{0j}(\vec{u},t)\right]$ by writing $P^i$ itself in terms of an integral, as
\begin{align}
     \left[P^i,T^{0j}(\vec{u},t)\right]&=  \int dA' dA \cdot \left[ T^{0i}(\vec{u}',t),   T^{0j}(\vec{u},t)\right] ,
\end{align}
where $dA'$ and $dA$ are infinitesimal area elements.

We can then write the commutator $ \left[ T^{0i}(\vec{u}',t),  T^{0j}(\vec{u},t)\right]$ in terms of a third, out-of-plane momentum density component $T^{0k}(\vec{u}',t)$ via a trivial algebra with structure constant $C^{full}$,
\begin{align}
     \left[ T^{0i}(\vec{u}',t),  T^{0j}(\vec{u},t)\right] = C^{k,full}_{ij}(\vec{u},t) T^{0k}(\vec{u},t)
\end{align}
We now take into account projection to the LLL in the sense that the momentum density operators $\{T^{0i} \}$ reduce to the field strength components $\{F^{0i}\}$ due to quenching of kinetic energy. We do not, however, attempt to take into account the fuzzification due to LLL projection discussed in the previous section, an approximation which simplifies calculations while still providing some insight into a possible topological invariant on fuzzy spaces. In anticipation of a potential central extension, we denote a counterpart algebra of the field strength components as
\begin{align}
     &\left[ F^{0i}(\vec{u}',t),  F^{0j}(\vec{u},t)\right] \\ \nonumber
     &= C^{k,F}_{ij}(\vec{u},t) F^{0k} (\vec{u},t) \\ \nonumber
     &= \left(C^{k,full}_{ij}(\vec{u},t)+ C^{k,proj}_{ij}(\vec{u},t) \right)F^{0k} (\vec{u},t),
\end{align}
to explicitly identify a deviation in the algebra of the momentum density due to projection to the LLL as $C^{k,proj}_{ij}(\vec{u},t) $. 

From the earlier examples, however, we can explicitly evaluate the LHS as
\begin{align}
     &\int  dA' \cdot \left[ F^{0i}(\vec{u}',t),  F^{0j}(\vec{u},t)\right] \\ \nonumber
     &= i\partial_iF^{0j}(\vec{u},t)+ip_\phi \epsilon_{ij}\partial_i\partial_j\phi(\vec{u},t).
\end{align}
We may then identify $dA \cdot C^{k,full}_{ij}(\vec{u},t) T^{0k} (\vec{u},t)=i\partial_iF^{0j}(\vec{u},t)=i\partial_iT^{0j}(\vec{u},t)$, as this first term is expected without projection to the LLL. The second term can then be identified with the deviation in the algebra due to projection to the LLL,
or
\begin{align}
    dA \cdot C^{k,proj}_{ij}(\vec{u},t) F^{0k} (\vec{u},t) = ip_\phi \epsilon_{ij}\partial_i\partial_j\phi(\vec{u},t),
\end{align} 
which ultimately yields the central extension in the case of the quantum ferromagnet and the Landau model. That is,

\begin{align}
\int dA \cdot C^{k,proj}_{ij}(\vec{u},t) F^{0k} (\vec{u},t) = 4 \pi \mc{Q}.
\end{align}
We can furthermore define $ \tilde{C}^{k,proj}_{ij} =  \left[ F^{0i}(\vec{u}',t),  F^{0j}(\vec{u},t)\right]\left(F^{0k} (\vec{u},t)\right)^{-1}-C^{k,full}_{ij}$ to compute $\mc{Q}$ as
\begin{align}
\int dA \cdot \tilde{C}^{k,proj}_{ij}(\vec{u},t) =4 \pi \mc{Q}.
\end{align}

This reformulation of the central extension illustrates how the non-trivial topological invariant associated with projection to the LLL can be interpreted as an algebraic artifact of the field strength algebra of an occupied subspace, in direct analogy to a non-trivial Chern number occurring as an algebraic artifact of projection to the occupied subspace~\cite{parameswaran2012}. This picture, while explicitly taking into account projection as changes in the forms of the momentum density operators, yet neglecting non-commutativity to realise a quantised topological invariant by integration, motivates the ansatz for the topological invariant on the fuzzy sphere in the following subsection as a non-trivial artifact of a Lie algebra associated with projection \textit{of operators for particular pspin observables} to an occupied subspace, given that such a formulation holds in the non-fuzzy limit.

\subsubsection{Computing topological invariants on severely-fuzzified spaces or for extra severely-fuzzified dimensions}
Here, we consider approaches to computing topological invariants over severely-fuzzified spaces or for systems with extra severely fuzzified dimensions. While the machinery  to write the Lagrange density is well-established~\cite{aschieri2007} and effectively the same between weak and severe fuzzification, topological characterization over spaces such as $M^{2+1} \times \left(S/R \right)_F$ is less developed in the physics literature, to our knowledge.  Past work, as mentioned in Section II, has attempted to compute topological invariants on weakly-fuzzified spaces in analogy to computation on non-fuzzy spaces~\cite{balachandran2006}. However, this approach has, to our knowledge, not yielded quantised values for topological invariants, which suggests a different, currently undefined, approach is required. The present work serves as motivation to develop or apply methods for computing topological invariants in these cases.  

In an effort to address this issue, we outline one possible method for computing quantised topological invariants on fuzzy spaces, making use of the interpretation of fuzzy coset space coordinates as isospin DOFs. This method is motivated by phenomenology and numerical results on the MCI in particular~\cite{banerjee2024}. In effect, given these isospin DOFs, we enlarge the local Hilbert space based on the remaining Cartesian spatial coordinates, and project the isospin operators to the occupied subspace. We denote the spin representation over occupied states as $S^i_{occ} = \rho_{occ}S_i$, in terms of the density matrix $\rho_{occ}$ and $S^i$ the pspin operator determined by the basis. The spin expectation value for the occupied states is then $\langle S_i \rangle = \tr \left[ \rho_{occ} S_i \right] = \tr \left[ S^i_{occ} \right]$. 

This interpretation of the fuzzy coset space coordinates as isospin DOFs corresponds to the gauge field components with fuzzy coordinates, e.g.,  $A_{a}$, reducing under severe fuzzification to a spin/position operator of the occupied states $S^{a}_{occ}$ itself in part as,
\begin{align}
 A_{a}&=-i\bar{\Psi} \left[ S_{a}, \Psi\right] \\ \nonumber
&= -iS^{a}_{occ} +i\bar{\Psi} \Psi S_{a} \nonumber \\
&= -iS^{a}_{occ} +iS_{a}.
\end{align}
Examining the field strength component over the fuzzy coset space coordinates,
\begin{align}
 F_{ab} &= \left[ S_{a}, A_{b} \right]   - \left[ S_b, A_a \right] + \left[A_a, A_b \right] - C^c_{ab} A_c,
 \end{align}
we enforce the equation of motion (EOM) constraint $F_{ab}=0$, corresponding to flat connection, to determine the \textit{structure factor} $C^{c}_{ab, occ}$, the generalisation of structure constant $C^{c}_{ab}$ after projection to the occupied subspace. Terms with position operators $S_{a}$ are eliminated, meaning only $\left[A_a,A_{b} \right]$ survives to give
  \begin{align}
  C^{c}_{ab,occ} = \left[ -iS^a_{occ}, -iS^b_{occ} \right](-iS^c_{occ})^{-1}.
 \end{align}
 Encoding the full structure factor as $\boldsymbol{C}_{EOM}$, the associated topological charge is then
 \begin{align}
    q_{EOM} &= k\tr\left[ \boldsymbol{C}_{EOM} \right],
\end{align}
 
We may generalise this further, by considering field strength components $F_{ab,s}$ as
\begin{align}
 F_{ab,s} &= \left[ S_{a}, A^b\right]   - \left[ S_{b}, A^a \right] + \left[A^a, A^b \right] - C^{c}_{ab,occ} A^c.
 \end{align}
The expected expression for topological charge $q$ of a LL\textsubscript{F} more generally is then,
\begin{align}
    q &= k\tr\left[ \boldsymbol{C} \right] \nonumber \\
    &= k\tr\left[\left[F_{ab,s}, F_{bc,s} \right] F^{-1}_{ca,s}  \right],
    \label{fuzzyq}
\end{align}
which is the fuzzified counterpart of the topological charge $q$ of a skyrmion in a NL$\sigma$M appearing as central extension of a momentum commutator~\cite{watanabe2014}. For the next section, it is useful to note that this algebra corresponds, in general, to some topological deformation of a fuzzy two-sphere, which we term the fuzzy (topological) two-sphere, or $S^F_{(T)F}$.

Banerjee~\emph{et al.}~\cite{banerjee2024} study a topological invariant of the type in Eq.~\ref{fuzzyq} numerically, finding the topological response signature of the MCI shows some evidence of quantisation of the expression to a value deviating from that of the unprojected spin representation for the system by a rational number. This motivates further investigation of Eq.~\ref{fuzzyq} and related expressions analytically and numerically to understand their potential as topological invariants within the QSkHE framework. For the time being, we employ this ansatz topological invariant in the next section to characterise the relationship between the skyrmion number $\mathcal{Q}$ defined in Section I, which characterises topologically non-trivial isospin expectation value textures over the BZ, and the second Chern number $C_2$.

\subsubsection{Anisotropic fuzzification of the four-sphere $S^4_F$}

In this section, we introduce methods for defining gauge theories over spaces with a mixture of Cartesian and fuzzy coset space coordinates, corresponding to space manifolds without \textit{global} product structure---as in the case of gauge theories with extra fuzzy dimensions---but still possessing \textit{local} product structure. We consider the Lagrange density defined over the manifold $M^{1+4}$, with four Cartesian space coordinates defining the four-sphere $S^4$. 

While one can naively fuzzify two Cartesian coordinates, it is unclear as to which manifold this yields, when starting with $M^{1+4}$. We circumvent this issue by taking advantage of the much simpler structure of the fuzzy four-sphere $S^4_F$. Higher-dimensional fuzzy spheres can be defined as a nested hierarchy of lower-dimensional fuzzy spheres~\cite{Hasebe:2010vp}. We formulate the action taking advantage of this, corresponding to first defining a fuzzy two-sphere $S^2_F$ with a finite number of fluxes. Construction of $S^4_F$ involves generalising each flux over $S^2_F$ itself to a fuzzy two-sphere $S^2_F$~\cite{Hasebe:2010vp}. We differentiate the two types of $S^2_F$ as $S^2_{F,xy}$ and $S^2_{F,zw}$. 

Further insight into this process can be gained by first considering a four-sphere $S^4$, which is defined for an EFT of interest in terms of the second Hopf map, as in Eq.~\ref{2ndHopfmap}. As in Bernevig~\emph{et al.}~\cite{bernevig6Dfieldtheory}, however, we may write the spinor $\Psi$ of Eq.~\ref{2ndHopfmap} in a form that more explicitly encodes the dimensional hierarchy of $S^4$, by writing it in terms of spinors defined in terms of first Hopf maps. That is, one may define two first Hopf maps in terms of NL$\sigma$M fields $\{n_{\alpha}\}$, with $\alpha \in \{xy,zw\}$, as
\begin{align}
    {n_{\alpha, i} \over R_{\alpha}} &= \bar{u}_{\alpha} \sigma_{\alpha, i} u_{\alpha},
\label{1stHopfmap}
\end{align}
where $R_{\alpha}$ is the radius of the $\alpha$ two-sphere $S^2_{\alpha}$ satisfying $\sum_i n_{\alpha, i}n^{\alpha,i} = R^2_{\alpha}$, $u_{\alpha}$ is a two-component spinor, and $\{\sigma_{\alpha,i}\}$ is a set of Pauli matrices defining the $\alpha$ first Hopf map. 

To fuzzify $S^4$, we place a monopole charge inside $S^4$. Taking into account the dimensional hierarchy of $S^4$, however, we can consider placing the monopole charge inside, for instance, $S^2_{zw}$ in the hierarchy. Projecting to the LLL, we fuzzify $S^2_{zw}$ to $S^2_{zw, F}$. As $S^2_{zw, F}$ serves as a generalised flux for $S^2_{xy}$, however, $S^2_{xy}$ is also fuzzified to $S^2_{xy,F}$, such that this placement of monopole charge fuzzifies the entire $S^4$ to $S^4_F$. 

There is then a total number of fluxes defining $S^4_F$, which must be related to the number of fluxes defining $S^2_{xy,F}$ and  $S^2_{zw,F}$ specifically. As $R_{xy}$ and $R_{zw}$ are potentially distinct, we may then fuzzify $S^4_F$ \textit{anisotropically}, rather than isotropically: we define position coordinates for 
$S^2_{F,xy}$ in terms of SU(2) generators with $N_{xy} \times N_{xy} $ matrix representation size, and position coordinates for 
$S^2_{F,zw}$ in terms of SU(2) generators with $N_{zw} \times N_{zw} $ matrix representation size. We note, however, that the $N_{xy}, N_{zw}$ are dependent, in the sense that the flux through $S^4_F$ is quantised and even after choosing the anisotropy we assume that it is still preserved.  To model the 2+1 D QSkHE, we take $N_{xy}\gg N_{zw}$, corresponding to severely fuzzifying $S^2_{F,zw}$ and weakly fuzzifying $S^2_{F,xy}$.

As outlined in the previous section, a dual interpretation of coset space position coordinates as isospin DOFs, rather than choosing one of these interpretations, whether a sphere is weakly or severely-fuzzified, is useful in defining quantised topological invariants for fuzzy coset spaces. If the non-commuting position coordinates of $S^2_F$ are projected to an occupied subspace, such as via projection to the LLL, the fuzzy two-sphere generalises to a (topological) fuzzy two-sphere $S^2_{(T)F}$, with potentially generalised algebraic structure encoded in a structure factor $C^{c,occ}_{a,b}$, with elements that are fuzzy counterparts of the Berry curvature. A trace applied to $C^{c,occ}_{a,b}$ can yield non-trivial topological charge $q$, generalising the Chern number computed by integrating Berry curvature.

We take advantage of the following results
\begin{enumerate}
\item The fuzzy sphere may be generalised to the fuzzy (topological) two sphere $S^2_{(T)F}$ to compute quantised topological invariants, overcoming a long-standing issue in topological characterisation of field theories defined over fuzzy coset spaces.
\item The fuzzy four sphere $S^4_F$ can be constructed by nesting fuzzy two-spheres $S^2_F$.
\item Fuzzy coset space position coordinates have dual interpretation as an isospin DOF, meaning they can enlarge a local Hilbert space. This dual interpretation is actually consistent with the dimensional hierarchy of the higher-dimensional fuzzy spheres.
\item The nested fuzzy two-spheres at different levels in the hierarchical structure comprising the fuzzy four-sphere may be fuzzified to different degrees.
\item The fluxes on the fuzzy two-sphere $S^2_F$, which are generalised to fuzzy two-spheres themselves to construct the fuzzy four-sphere $S^4_F$, may each more generally be interpreted as a local field strength encoded as an element of the field strength matrix $F_{xy}$. The corresponding fuzzy two-sphere generalisation of each flux is then generalised to a fuzzy (topological) two sphere $S^2_{(T)F,zw}$, meaning the element of $F_{xy}$ is generalised to the structure factor matrix $C^{c,occ}_{zw}$ of $S^2_{(T)F,zw}$.
\end{enumerate}

The counterpart of the CS action on $M^{1+4}$, for the case of $M^1 \times S^4_F$, can then be written in terms of a CS Lagrange density matrix $\mathcal{L}_{F}$ for the fuzzy four-sphere~\cite{Alekseev:2000fd}, with $\left(N_{xy}N_{zw}\right)^2$ entries, as

\begin{align}
    S = \int dt k_{xy}\tr\left[k_{zw}\tr\left[ \mathcal{L}_{F}\right] \right],
\end{align}

where  $k_{z w}\tr$ and $k_{x y}\tr$ correspond to implicitly including $zw$ and $xy$ dependence in the Lagrange density $\mathcal{L}_{F}$. That is, the $x_0$\textsuperscript{th} row and $y_0$\textsuperscript{th} column of $\mathcal{L}_{F}$ prior to introducing $k_{z w}\tr$ contains the three-form entry of matrix $\hat{f}_{x_0 y_0}$, and $k_{z w}\tr$ implies generalisation from $\hat{f}_{x_0 y_0}$ at location $\left( x_0, y_0\right)$ on $S^2_{(T)F,xy}$ to the matrix $\hat{f}_{x_0 y_0}\wedge \hat{F}_{x_0 y_0}$, where $\hat{F}_{x_0 y_0}$ is the field strength over $S^2_{(T)F,zw}$ specific to position $\left(x_0, y_0 \right)$ on $S^2_{(T)F,xy}$. This approach employs the dimensional hierarchy of fuzzy spheres to treat the $z$ and $w$ coordinates as isospin DOFs at each site of $S^2_{(T)F,xy}$.

As a brief aside, since the four-sphere $S^4$ can similarly be interpreted as a nesting of two-spheres $S^2$ as stated earlier in this section, the Lagrange density of the 4+1 D SU(2) theory can similarly be constructed from a Lagrange density for a 2+1 D U(1) theory, generalised by promoting the Lagrange density at each point on $S^2$ itself to contain an instance of a Lagrange density for a second $S^2$. Compactly, we express this promotion of the Lagrange density corresponding to local product structure as
\begin{align}
dA \rightarrow dA \wedge dA,
\end{align}
and similarly for terms related by gauge invariance, corresponding to generalising from the spinor for the first Hopf map over the $S^2_{xy}$ at position $\left( x_0, y_0 \right)$ on the  $S^2_{xy}$ with  the spinor for the first Hopf map over the  $S^2_{zw}$ specific to $\left( x_0, y_0 \right)$. We note that this also suggests an alternative approach to computation of central extensions for higher-dimensional spheres $S^d$. Rather than deriving the central extension for the four-sphere $S^4$ simply by studying the Nambu geometry associated with the Dirac matrices satisfying the Clifford algebra~\cite{langmann2024}, it may instead be possible to derive the central extension can be derived from a nesting of commutators, one for each $S^2$ in the hierarchy corresponding to $S^4$, a point we will explore in future work.

Returning to the EFT with anisotropic fuzzification, we now discuss topological invariants. Computing the fuzzy counterpart of the second Chern number, $C_2$, over the fuzzy four-sphere then involves generalising from computation of the structure factor of $S^2_{(T)F,xy}$, $\boldsymbol{C}_{xy}$, to enlarging the element of $\boldsymbol{C}_{xy}$ for each $\left(x_0, y_0\right)$, $\boldsymbol{C}_{x_0y_0}$, to a matrix containing the structure factor of $S^2_{(T)F,zw}$ for that element, $\boldsymbol{C}_{x_0 y_0}^{zw}$. That is, 
\begin{align}
    \boldsymbol{C}_{x_0y_0} &\rightarrow \boldsymbol{C}_{x_0y_0} \otimes \boldsymbol{C}_{x_0 y_0}^{zw},
\end{align}
where $\otimes$ denotes Kronecker product.

We write the full, generalised structure factor more compactly as $\boldsymbol{C}_{xy}^{zw}$ and the second Chern number over the fuzzy (topological) four sphere, $c_2$, as
\begin{align}
    c_2 &= k_{xy}\tr \left[ k_{zw}\tr \left[ \boldsymbol{C}_{xy}^{zw}  \right] \right].
\end{align}

However, it's also useful to reinterpret $\boldsymbol{C}_{x_0y_0} \otimes \boldsymbol{C}_{x_0 y_0}^{zw}$ more explicitly in terms of isospin DOFs of  $S^2_{(T)F,xy}$. The tensor product introduces an isospin DOF with matrix Lie algebra generator three-form $\boldsymbol{S}$. We may then define projection of these generators to the local occupied subspace at $\left(x_0 y_0 \right)$ as $\boldsymbol{S}^{occ}_{x_0y_0} = \rho_{x_0y_0} \boldsymbol{S}$, where, $\rho_{x_0 y_0}$ is the component of the density matrix $\rho$ for entry $\left(x_0, y_0 \right)$. Thus  $\tr\left[\rho_{x_0y_0} \boldsymbol{S} \right] = \langle \boldsymbol{S}(x_0,y_0) \rangle$, the expectation value of $\boldsymbol{S}$ at site $\left(x_0,y_0\right)$. $ \boldsymbol{C}_{x_0y_0} \otimes \boldsymbol{C}_{x_0 y_0}^{zw}$ is then a fuzzy counterpart of the local skyrmion curvature~\cite{qskhe, cook2023}, or
\begin{align}
     \boldsymbol{C}_{x_0y_0}\otimes \boldsymbol{C}^{x_0 y_0}_{zw} &=\boldsymbol{\Omega}_S(x_0,y_0) \nonumber \\
     &= \left[ S^{occ}_{x}(x_0,y_0) , S^{occ}_{y}(x_0,y_0) \right] \left(S^{occ}_{z}(x_0,y_0)\right)^{-1}.
\end{align}
Taking $N_{xy}$ to be very large such that it is well-approximated as $\infty$ relative to $N_{zw}$, we write this in its non-fuzzy form as
\begin{align}
     \Omega_S(x,y) &= \boldsymbol{S}^{occ}(x,y) \cdot \left( \partial_{x}\boldsymbol{S}^{occ}(x,y) \times \partial_{y} \boldsymbol{S}^{occ}(x,y)  \right) 
\end{align}

Integrating over $x$ and $y$ assuming $t$ independence yields the spin skyrmion number of the topological skyrmion phases, or 
\begin{align}
    \mathcal{Q} = {1 \over 4 \pi}\int dxdy \Omega_S(x,y).
\end{align}
This reveals $\mathcal{Q}$ as the second Chern number of a severely ``squashed'' fuzzy four-sphere $S^4_{F}$, corresponding to $N_{xy}/N_{zw} \approx \infty$.

The second Chern number, then, may also be computed from a counterpart strictly 2+1 D gauge theory, which derives from the more general theory with severe fuzzification by tracing out the fuzzy coset space coordinates, in direct analogy to computation of observable-enriched entanglement introduced in Winter~\emph{et al.}~\cite{winterOEPT} from study of lattice models. $Q$ of the 2+1 D QSkHE can therefore be expressed---and computed---as a first Chern number associated with an effective gauge field and action for a spin subsystem~\cite{winterOEPT}, despite actually being a second Chern number in the simplest cases. Such a gauge theory only captures some of the phenomenology of the full action, however, due to effective reduction of LL\textsubscript{F}s to point objects carrying unit charge.

This procedure of anisotropic fuzzification, and the notion of the fuzzy (topological) sphere, motivates study of finite-size topological phases specifically associated with the 3+1 D Hopf insulator phase~\cite{moore2008}, furthermore, as part of efforts to apply fuzzification to reduce dimensionality by eliminating one Cartesian coordinate at a time. As the Hopf insulator is captured by a two-band Bloch Hamiltonian, opening boundary conditions in one direction and thinning the system to be two unit cells thick in this direction yields a four-band Bloch Hamiltonian with two Cartesian spatial coordinates, the case most closely-related to gauge theories of the second Hopf map studied here.  With appropriate symmetries, then, the thin film Hopf insulator is described by a field theory for the 2+1 D QSkHE derived by anisotropic fuzzification of the 4+1 D SU(2) gauge theory. In particular, given the notion of the fuzzy (topological) two-sphere $S^2_{(T)F}$, opening boundary conditions of the 3+1 D Hopf insulator in one direction introduces a pspin DOF and associated algebra, but also an algebra associated with projecting the pspin representation to an occupied subspace. From this second algebra, it may be possible to reverse the fuzzification to construct a 4+1 D Hamiltonian capturing the topology of the 3+1 D Hopf insulatorm, a direction of future work.

We note that interpretation of the four sphere $S^4$ in terms of nested two-spheres $S^2$ provides insight into the embarrassment of riches of the 4+1 D QHE~\cite{zhanghu2001}, which is the presence of high-helicity excitations: two of the spatial dimensions are necessarily also internal DOFs of the first $S^2$ in the hierarchical picture of $S^4$. Given this dual interpretation, the 4+1 D QHE is better understood as an unphysical precursor of the 2+1 D QSkHE. While the 4+1 D QHE is known to be important for its role as a parent state from which lower-dimensional topological states descend~\cite{qi2008TRIFT}, its deeper significance is as a precursor of the generalisation of the 2+1 D QHE to the 2+1 D QSkHE.

Interestingly, model Hamiltonians for the 4+1 D CI do not suffer from this problem of very high-helicity excitations of the 4+1 D QHE~\cite{qi2008TRIFT}. Some insight into the reason for this comes from examining the form of the Bloch Hamiltonian $H(\boldsymbol{k})$ carefully. We may write $H(\boldsymbol{k})$ in terms of a five-vector of momentum $\boldsymbol{k}$-dependent scalar functions and a vector of five mutually anticommuting Dirac matrices $\boldsymbol{\Gamma}$ as $H(\boldsymbol{k}) = \boldsymbol{d}(\boldsymbol{k}) \cdot \boldsymbol{\Gamma}$. We can then associate a four-sphere $S^4$ with the normalised $\boldsymbol{d}(\boldsymbol{k})$ vector itself, which may be defined by a map similarly to definition of non-linear sigma model (NL$\sigma$M) fields defined in terms of first or second Hopf maps~\cite{bernevig6Dfieldtheory}. In addition, however, the Dirac matrix vector $\boldsymbol{\Gamma}$ defines, in higher-symmetry cases, a severely-fuzzified two-sphere as $\left(SO(5)/SU(2) \right)_F \cong S^2_F$. This $S^2_F$ can serve as the internal DOFs for two separate two-spheres $S^2$ in a hierarchy encoded within the $\boldsymbol{d}(\boldsymbol{k})$-vector. That is, the pair of $S^2$'s can be linked by $S^2_F$, such that neither $S^2$ serves as the internal DOFs of the other $S^2$. Instead, $S^2_F$ encodes internal DOFS of each $S^2$, such that high-helicity excitations are not present as helicity is bounded by the representation size $N$ of generators for $S^2_F$, which is finite and furthermore severely-fuzzified ($N$ is small). As a result of this additional $S^2_F$, the 4+1 D CI is actually a special case of an underlying 6+1 D theory in the simplest cases.

We may similarly apply this reasoning to the two-band 2+1 D CI. Writing the Bloch Hamiltonian as $H(\boldsymbol{k}) = \boldsymbol{d}(\boldsymbol{k}) \cdot \boldsymbol{\sigma}$ in terms of a three-component $\boldsymbol{d}(\boldsymbol{k})$-vector of momentum-dependent scalar functions, and the vector of Pauli matrices $\boldsymbol{\sigma}$, we associate a two-sphere $S^2$ with the normalised $\boldsymbol{d}\left(\boldsymbol{k} \right)$-vector, and a severely-fuzzified two-sphere with the vector of Pauli matrices as $\left(SU(2)/U(1) \right)_F \cong S^2_F$. That is, the 2+1 D CI is a special case of an underlying 4+1 D theory. From this perspective and the connection between the 2+1 D CI and the quantum Hall ferromagnet/bilayer (QHFM)~\cite{PhysRevB.47.16419,girvin_multi}, we see the physics of these systems is more generally captured by the gauge theory of the 2+1 D QSkHE. This indicates the magnetic phenomena of the QHFM (and two-band CI) are a useful guide in understanding magnetic response signatures of the QSkHE, and a simplified case of this physics. 

 We briefly comment on magnetic response signatures of the QSkHE given these insights from the EFT, which is being explored in detail in a separate work. In these cases of the QHFM and the two-band CI, the spin topology is slaved to the charge topology, such that similar signatures appear in the spin susceptibility to those in the charge susceptibility~\cite{jiang2011}. In simpler cases, non-trivial skyrmion number $\mc{Q}$ and trivial Chern number in more complex systems are expected to yield similar spin susceptibility signatures of topology to those for the QHFM and two-band CI, \textit{without} corresponding features in charge susceptibility, and these spin signatures of the QHFM and two-band CI should not be interpreted as consequences necessarily of a non-trivial Chern number, but rather non-trivial skyrmion number. In more complex cases, however, magnetic response signatures are expected to become intrinsically higher-dimensional.

\subsubsection{Intertwined bulk-boundary correspondence and topological response signatures from the perspective of the thin brane limit of the 6+1 D QHE}

Here, we explore possible mechanisms for the intertwining of bulk-boundary correspondence and response signatures in lattice models, most notably the signatures in 2+1 D four-band Bloch Hamiltonian tight-binding models most closely associated with the topological response sigantures of the 4+1 D CI, such as the current density across the system in the direction of OBCs and associated boundary chiral anomalies. 

In the EFT of the 6+1 D QHE, membranes constructed from $\mathbb{C}P_3 \rightarrow S^4$ wrap a spherical $2$-cycle of $\mathbb{C}P_3$. They are characterised by the topological current~\cite{bernevig6Dfieldtheory},
\small
\begin{equation}
J^{\Gamma \Gamma_1 \Gamma_2} = {1 \over 4!} \epsilon^{\Gamma \Gamma_1 \Gamma_2 \Gamma_3 \Gamma_4 \Gamma_5 \Gamma_6} \epsilon_{abcde} X^a \partial_{\Gamma_3} X^b \partial_{\Gamma_4} X^c \partial_{\Gamma_5}X^d \partial_{\Gamma_6}X^e,
\end{equation}
\normalsize
where the $X^{a}$'s are $O(5)$ sigma model-like fields constructed from the embedding fields $\Psi(x)$. The membranes are non-trivial topological configurations of $O(5)$ sigma-model fields, which wind around a spherical $4$-cycle in $\mathbb{C}P_3$, stabilized by homotopy group $\pi_4(S^4) = \mathbb{Z}$. The conserved topological charges are given by 
\begin{equation}
    \mathcal{Q}^{\Gamma_1 \Gamma_2} = {1 \over \Omega_4} \int_{S^4} J^{0 \Gamma_1 \Gamma_2},
\end{equation}
where $\Omega_4$ is the volume of the $S^4$ manifold.

We can also construct $4$-branes in the fluid, which wrap spherical $4$-cycles of $\mathbb{C}P_3$ and are constructed as maps $\mathbb{C}P_3 \rightarrow S^2$. They are characterised by the topological current
\begin{equation}
J^{\Gamma \Gamma_1 \Gamma_2 \Gamma_3 \Gamma_4} = {1 \over 2} \epsilon^{\Gamma \Gamma_1 \Gamma_2 \Gamma_3 \Gamma_4 \Gamma_5 \Gamma_6} \epsilon_{ijk} n^i \partial_{\Gamma_5} n^j \partial_{\Gamma_6} n^k.
\end{equation}
Here, the $n^i(x)$'s are $\mathrm{O}(3)$ sigma model-like fields given by $n^i(x)/r = \bar{u}(x) \sigma^i u(x)$, with 
\begin{equation}
    u = \begin{pmatrix}
        u_1 \\
        u_2
    \end{pmatrix} = \begin{pmatrix}
        \sqrt{{r+n_3\over 2r}}\\
        {n_1 + i n_2 \over \sqrt{2r (r+n_3)} }
    \end{pmatrix}
\end{equation}
defined by the first Hopf map
\begin{equation}
    {n_i \over r} = \bar{u} \sigma_i u, \hspace{3mm} i=1,2,3.
\end{equation}
The $4$-branes are non-trivial topological configurations of $\mathrm{O}(3)$ sigma model fields winding a spherical $2$-cycle of $\mathbb{C}P_3$. They are stabilised by homotopy group $\pi_2(S^2) = \mathbb{Z}$ and the conserved topological charges are given by
\begin{equation}
    \mathcal{Q}^{\Gamma_1 \Gamma_2 \Gamma_3 \Gamma_4} = {1 \over \Omega_2} \int_{S^2} J^{0 \Gamma_1 \Gamma_2 \Gamma_3 \Gamma_4},
\end{equation}
where $\Omega_2$ is the volume of the $S^2$ manifold.

In the limit of thin objects, the single $2$-brane or $4$-brane Lagrangians are generalisations of the single-particle Lagrangian. The higher form gauge fields felt by the extended objects can be constructed explicitly out of the $1$-form background gauge field $A$~\cite{bernevig6Dfieldtheory} as 
\begin{equation}
    C_3 = A \wedge dA, \hspace{3mm} C_5 = {1 \over 2} A \wedge dA \wedge dA.
\end{equation}
The $4$-brane and $2$-brane move in the $X^a$ directions perpendicular to the relevant effective magnetic field associated to the $3$- or $5$-form, respectively. That is, the branes behave effectively like particles in the $X^a$ space. 

In the thin brane regime, we can therefore decompose the 6+1 D system into a four-brane which sees an effective membrane magnetic field. In this regime, the four-brane responds to the membrane field similarly to a particle. If we consider generalized compactification of the present work for two coordinates of the four-brane and the two coordinates of the membrane, we arrive at a 6+1 D theory compactified to 2+1 D, which contains intrinsically 4+1 D topology of the four-brane, plus an effective external field corresponding to the compactified membrane. 

Within this framework, we can understand topological response signatures even in the absence of external fields, as discussed in Section I, by generalising from the thin brane limit. In this regime, the four-brane providing effective external fields is  entangled with the two-brane, meaning the two-brane can effectively provide external fields to itself. The cases of high-symmetry four-band models are potentially more subtle---though similar---to the above argument. In the 5+1 D U(1) $\times$ U(1) multiplicative gauge theory, breaking the tensor product structure of the spinor---while respecting appropriate symmetries---to realize the SU(2) gauge theory entangles the parent pspin DOF sectors to yield a similar, yet more constrained, intertwining of bulk-boundary correspondence and topological response signatures.

\section{Discussion \& Conclusion}

In this work, we introduce the effective field theory (EFT) of the quantum skyrmion Hall effect (QSkHE), a generalisation of the quantum Hall effect (QHE) framework, in which intrinsically $\delta+1$-dimensional ($\delta$+1 D)---with $\delta>0$---compactified many-body states defined over severely-fuzzified manifolds can play the role, in the QSkHE, that charged particles play in the QHE. This is consistent with the interpretation that (pseudo)spin (pspin) degrees of freedom (DOFs), where pseudo indicates the spin could correspond to myriad physical quantities, associated with matrix representation size of associated Lie algebra generators $N \times N$, encode $\delta$ spatial dimensions both when $N$ is large, as considered previously in study of non-commutative field theories~\cite{susskind2001_qh, SimeonHellerman_2001, gavriil2015higher,aschieri2007, Aschieri:2003vy,Aschieri:2004vh}, but also when $N$ is small and the pspin DOF is clearly interpreted as an isospin, or internal DOF. This EFT is developed to explain phenomenology of the QSkHE identified in separate works~\cite{qskhe, cook_multiplicative_2022, cook2023, cookFST2023}, which indicate topological states of myriad isospin DOFs in systems with $d$ Cartesian space coordinates exhibit bulk-boundary correspondence and response signatures identifiable with those of previously-known $d$ +$\delta$+$1$ D topologically non-trivial phases of matter.

We construct the EFT of the QSkHE first by considering a minimal EFT of the multiplicative Chern insulator (MCI). Given the $4 \pi$ Aharonov-Bohm effect of the MCI~\cite{banerjee2024}, we construct this EFT as a minimal $K$-matrix CS theory, focusing specifically on realisation of a $\nu=1/2$ FQH plateau to support numerics on the MCI~\cite{banerjee2024}. We note that, even at this level of treatment, the EFT illustrates the importance of the multiplicative topological phases of matter---and the QSkHE more generally---for topological quantum computation.

Motivated by phenomenology of lattice tight-binding models realising topologically non-trivial phases of matter associated with the QSkHE, which possess two Cartesian space coordinates ($D=2$) previously-related to the 2+1 D SU(2) gauge theory, we treat the isospin DOF of these models within the framework of extra fuzzy dimensions~\cite{aschieri2007, Aschieri:2003vy, Aschieri:2004vh, gavriil2015higher}. We therefore focus on scenarios where the $\delta$+1 D severely-fuzzified many-body state is a severely-fuzzified Landau level (LL\textsubscript{F}), with some number of filled orbitals. We furthermore employ the dimensional hierarchy of (fuzzy) higher-dimensional spheres~\cite{Hasebe:2010vp} to generalise the extra fuzzy dimensions framework to the case of space manifolds with local product structure to show the spin skyrmion number  $\mathcal{Q}$ of topological skyrmion phases is a second Chern number $C_2$ of a ``severely-squashed'' fuzzy (topological) four-sphere $S^4_{(T)F}$. These efforts motivate deeper investigation of topology on non-commutative spaces, especially in cases of severe fuzzification.

The Lagrange density for the 2+1 D SU(2) gauge theory considered previously in the literature~\cite{Demler:1998pm, affleck1988, lee1998} is therefore a special case of a more general Lagrange density possible for the 2+1 D SU(2) gauge theory, which includes terms which are intrinsically $d$+1 dimensional, where $d\leq 4 $, with terms of dimensionality $d>2$  compactified---but still relevant---to the ostensibly 2+1 D gauge theory. Given the importance of these additional terms in defining compactified many-body states that include pairing between the parent DOFs suggested by the $4 \pi$ Aharonov Bohm effect of the MCI~\cite{banerjee2024}, and the importance of the 2+1 D SU(2) gauge theory in study of resonating valence bond states \cite{ANDERSON1973153,PhysRevB.35.8865}, the quantum spin liquid state~\cite{ANDERSON1973153, ANDERSON87, BASKARAN1987973, PhysRevLett.59.2095, PhysRevB.35.8865, PhysRevLett.61.2376, PhysRevB.37.3774, PhysRevLett.66.1773, PhysRevB.40.7387, PhysRevB.44.2664, PhysRevB.62.7850, PhysRevLett.86.292, wegner1971duality, RevModPhys.51.659, PhysRevLett.86.1881, KITAEV20032, KITAEV20062}, non-Fermi liquid physics\cite{PhysRevB.14.1165, PhysRevB.48.7183, RevModPhys.79.1015, RevModPhys.73.797, PhysRevB.78.035103, PhysRevB.8.2649, PhysRevB.40.11571, PhysRevB.50.14048, PhysRevB.50.17917, polchinski1994low, PhysRevLett.63.680, PhysRevB.46.5621, PhysRevB.46.5621, nayak1994non, hartnoll2018holographic}, and unconventional superconductivity~\cite{Demler:1998pm, PhysRevB.38.2926, lee1998, sachdev2018topological, PhysRevX.8.021048, PhysRevLett.119.227002, sachdev2016novel}, our results are potentially fundamental to understanding of these phenomena. It is therefore important to explore the implications of the generalised SU(2) theory introduced in this work for this physics. We 
 furthermore emphasize that the generalized compactification discussed in the present work is the minimal generalisation of quantum field theories required for consistency with the QSkHE and related topological states~\cite{qskhe}. This work therefore provides guidance particularly to high-energy physics as to which compactification schemes are physically-relevant. We also note that rich exchange statistics are possible for branes~\cite{Hasebe:2014nia, bernevig6Dfieldtheory}, indicating counterpart compactified many-body states of the QSkHE could greatly enrich quantum computation schemes even in systems with just two Cartesian spatial coordinates.

Notably, this EFT of the QSkHE provides insight into previously-unexplained experimental results on HgTe quantum wells (QWs) characterised by the Bernevig-Hughes-Zhang (BHZ) model~\cite{QSHI-HgTe-Theory}, revealing them as consistent with the QSkHE and its EFT~\cite{Demler:1998pm, bernevig6Dfieldtheory, qi2008TRIFT}. Related work by Ay~\emph{et al.}~\cite{ay2024} is reviewed here as evidence of the need to generalise the 2+1 D SU(2) gauge theory given discovery of the QSkHE, and provides strong evidence that past experimental work on HgTe quantum wells (QWs) is potentially the first known experimental observation of signatures of the QSkHE~\cite{ma_unexpected_2015} beyond the simpler framework of the QHE. These works therefore serve as a call for careful scrutiny of HgTe QW experiments specifically for further evidence of the QSkHE.

An important note concerning the EFT of the QSkHE in 2+1 D treated as a compactified  EFT of the 4+1 D QHE, specifically, is also that this resolves a long-standing issue of the 4+1 D QHE, that of the embarrassment of riches associated with high-spin particles~\cite{zhanghu2001}. In essence, if the dimensional hierarchy of the four-sphere is taken into account, the coordinates of the second two-sphere in the hierarchy are necessarily interpreted as internal DOFs of the first two-sphere in the hierarchy, meaning the isospin DOF is locked, in a sense, to two Cartesian coordinates and therefore yields unphysically-large helicity excitations. Compactification involves restricting to small spin values, thereby resolving this issue. 

In this sense, the effective field theories of higher-dimensional QHEs are more meaningful after generalized fuzzification to the effective field theories of the QSkHE. Related to this point, we note that the 4+1 D CI does not suffer from this issue of unphysically high helicity, and provides insight into how to avoid it. The Bloch Hamiltonian of the 4+1 D CI may be expressed as $\boldsymbol{d} \cdot \boldsymbol{\Gamma}$, where the $\boldsymbol{d}$-vector itself is defined in terms of the second Hopf map $S^7 \rightarrow S^4$ and encodes a four-sphere $S^4$. In addition, however, the Dirac matrices $\left\{\Gamma_i \right\}$ define a fuzzy coset space isomorphic to the fuzzy two sphere, or $\left(SO(5)/SU(2) \right)_F \cong S^2_F$ in higher-symmetry cases. This fuzzy two-sphere \textit{is in addition to the four sphere $S^4$}, and can encode the isospin DOFs of the two two-spheres $S^2$ in the dimensional hierarchy of the four-sphere $S^4$. As the isospin DOF is now encoded in coordinates of a severely-fuzzified sphere rather than Cartesian coordinates, the maximum helicity is no longer unphysically large. This furthermore implies that the 4+1 D CI is a lattice counterpart and very special case of the 6+1 D QHE in high-symmetry cases, rather than the 4+1 D QHE.

Similarly, the two-band CI with Bloch Hamiltonian $\boldsymbol{d} \cdot \boldsymbol{\sigma}$ encodes a two-sphere $S^2$ via the first Hopf map associated with $S^3 \rightarrow S^2$ via the $\boldsymbol{d}$-vector, and the vector of Pauli matrices $\boldsymbol{\sigma}$ encodes at additional fuzzy coset space $\left( SU(2)/U(1)\right)_F \cong S^2_F$, meaning the two-band CI---and the related field theory of the quantum Hall ferromagnet  (QHFM)~\cite{yang1994, karlhede1996, PhysRevLett.73.874, EOM00, PhysRevB.49.17208, PhysRevB.51.14725, KASNER1995289, PhysRevB.54.R2331, PhysRevLett.76.3204, V_N_Nicopoulos_1995}---is actually a very special case of the 2+1 D QSkHE where the pspin topology is slaved to the charge topology. That is, various magnetic response signatures of the QHFM should be interpreted as a very special case of magnetic response signatures of the QSkHE, rather than topology of the QHE framework. That is, these magnetic signatures are due to two topological invariants, the Chern number and the skyrmion number $\mathcal{Q}$, being locked in value to one another, rather than being consequences purely of non-trivial Chern number.

We close the present work by returning to answer the question with which we began, concerning the interpretation of spin angular momentum encoded in a matrix Lie algebra with generators of matrix representation size $N \times N$, where $N$ is potentially small. This is the regime where spin has previously been treated as a label, rather than as encoding spatial dimensions as in more recent works on non-commutative field theories~\cite{susskind2001_qh, SimeonHellerman_2001, gavriil2015higher,aschieri2007, Aschieri:2003vy,Aschieri:2004vh}. The QSkHE and its EFT reveal spin angular momentum encodes some finite number of spatial dimensions even for small $N$, in general, contrary to past theoretical treatments, in the sense that spin angular momentum can encode an intrinsically $\delta$+$1$ D topologically non-trivial many-body state in this regime of small $N$. Such many-body states can furthermore play the role in the QSkHE that charged particles play in the QHE, yielding intrinsically $d$+$\delta$+1 D states in systems with $d$ Cartesian space coordinates as well as pspin DOFs encoding $\delta$ dimensions in fuzzy coset space coordinates. This motivates review of myriad problems of quantum mechanics involving spin angular momentum encoded in matrix Lie algebras, and reveals the significant impact discovery of the QSkHE and this interpretation of spin could have on quantum information physics and related problems in many areas of modern physics.

\bibliography{main.bib}

\end{document}